\documentstyle[psfig]{mn2e}

\newcommand{\etal}{et al.}

\newcommand{\Ein}{{\em Einstein}}

\newcommand{\Ros}{{\em ROSAT}}
\newcommand{\ltsim}{\raisebox{-1mm}{$\stackrel{<}{\sim}$}}

\newcommand{\eg}{e.g. }
\newcommand{\ie}{i.e. }
\newcommand{\etc}{etc.}
\def\arcm{\hbox{$^\prime$}}

\title[ROSAT PSPC observations of nearby spiral galaxies - II. 
Statistical properties]
{ROSAT PSPC observations of nearby spiral galaxies - II. Statistical properties}
\author[A. M. Read \& T. J. Ponman]
 {A. M. Read$^{1,2}$ \& T. J. Ponman$^{2}$ \\
$^{1}$Max-Planck-Institut f\"ur extraterrestrische Physik,
Gie\ss enbachstra\ss e, D--85748 Garching, Germany \\
$^{2}$School of Physics and Space
Research, University of Birmingham, Edgbaston, BIRMINGHAM, B15 2TT 
}

   \date{Received date; accepted date}

\begin{document}

\label{firstpage}

\maketitle

\begin{abstract}
We present a statistical analysis of the largest
X-ray survey of nearby spiral galaxies in which diffuse emission
has been separated from discrete source contributions. 
Regression and rank-order correlation analyses are used to compare
X-ray properties such as total, source and diffuse luminosities, and
diffuse emission temperature, with a variety of physical and
multi-wavelength properties, such as galaxy mass, type and activity,
and optical and infrared luminosity.

The results are discussed in terms of the way in which hot gas
and discrete X-ray sources scale with the mass and activity of galaxies, 
and with the star formation rate. We find that the X-ray properties
of starburst galaxies are dependent primarily on their star-forming 
activity, whilst for more quiescent galaxies, galaxy mass is the
more important parameter. One of the most intriguing results 
is the tight linear scaling between far-infrared and diffuse X-ray luminosity
across the sample, even though the hot gas changes from a hydrostatic corona
to a free wind
across the activity range sampled here. 


\end{abstract}


\section{Introduction}
\label{sec_intro}

Though it was the \Ein\ satellite that really began the study of the X-ray
emission from nearby galaxies, its instrumentation only enabled detailed
studies of the very closest systems (\eg\ M31, M33 and the Magellanic Clouds).
It was only with the launch of 
\Ros\ (Tr\"{u}mper 1992) that detailed work on more distant systems,
or any reliable spectral work, became possible.

Results from both \Ein\ (see \eg\ the X-ray catalogue of Fabbiano, Kim \&
Trinchieri (1992)) and \Ros\ (\eg\ Read, Ponman \& Strickland 1997) 
suggest that normal spiral galaxies emit
X-rays in the soft X-ray (0.1$-$2.4\,keV) band with luminosities of
$\sim10^{38}-10^{42}$\,erg s$^{-1}$. This emission originates from a combination of
the integrated output of supernova remnants and accreting close binaries (low-
and high-mass X-ray binaries, cataclysmic variables \etc), and hot gaseous
diffuse emission. The latter may take the form of hot outflows (or winds), as seen 
emanating from the discs of nearby starburst galaxies (Watson, Stanger
\& Griffiths 1984, Fabbiano 1988), or hot coronal phases of the ISM, seen
within more normal galaxies (\eg\ Bregman \& Pildis 1994; Wang \etal\ 1995). 
So, the X-ray emission from spiral galaxies is generally composite and
complex. Hence simple integrated analyses, which combine the different
components and extract mean parameters for the whole, are of rather limited
value.

In Read, Ponman \& Strickland (1997), hereafter Paper~1, we reported
the results of the largest X-ray survey of nearby spiral
galaxies performed with the \Ros\ PSPC which involves a careful
separation of diffuse and point-like emission -- a uniform analysis of a sample of
17 nearby spiral systems covering a wide range in activity and inclination. 
The separation of point sources from the diffuse emission
allows an insight into the properties of X-ray sources within external
galaxies, and enabled us to establish the prevalence and properties of hot
galactic winds and halos. We discussed the X-ray properties of both the
point-source emission and the diffuse emission within each of the 17 systems
in detail, together with the general X-ray properties of the whole survey.
Physical parameters for the hot gas observed within many of these systems
were also derived.

We concluded from this study that nearby non-active spiral galaxies have
(0.1$-$2.0\,keV) X-ray luminosities in the range $10^{38}$-$10^{41}$\,erg
s$^{-1}$ and that significant amounts (up to $10^{9}$\,M$_{\odot}$) of
diffuse hot ($1-8\times10^{6}$\,K) gas are evident in most of
the systems observed (the amount of gas appearing to increase with $L_{X}$).
This hot gas extends above the planes of high-activity galaxies in
the form of winds and halos. Point sources within these galaxies 
were found to have (0.1$-$2.0\,keV) X-ray luminosities in the range of a few
$\times10^{35}$ to a few $\times10^{39}$\,erg s$^{-1}$, and dominate
the emission from the low-activity systems. Most of the sources with $L_{X} >
10^{39}$\,erg s$^{-1}$ were nuclear sources with soft ($kT \ltsim
1$\,keV) spectra.

In recent years, several related studies have appeared. The most similar
to our own work is the X-ray mini-survey of nearby edge-on spiral galaxies 
by Dahlem, Weaver \& Heckman (1998). This study has several galaxies
in common with our sample, and there is in general 
good agreement between the two studies for these systems (note that the 
Dahlem \etal\ (1998) study makes use also of {\em ASCA} data and more complex
spectral fitting). 
Other recent galaxy surveys using \Ros\ data include
a multivariate statistical analysis of integrated spiral galaxy 
luminosities (Fabbiano \& Shapley 2000), and a \Ros\ HRI survey of bright 
nearby galaxies (Roberts \& Warwick 2000) focusing on the point source
content.

In the present paper we use the results presented in Paper~1 to analyse the
statistical properties of the sample as a whole. This study is the best of its
kind to date, in terms of the size of the sample coupled with the detail of
the X-ray analysis. The way in which the X-ray properties of these spiral
galaxies, such as their X-ray luminosity, their diffuse fraction and the
temperature of the diffuse emission, vary with other properties, such as
galaxy morphology, activity and optical and infrared luminosity, are discussed.

The plan of this paper is as follows. After some brief comments
(Section~\ref{sec_sample}) on the selection of the original sample, the
observations and the data reduction methods used, the results of statistical
properties of the sample as a whole are presented (Section~\ref{sec_results}).  
Important implications of these results are discussed in 
Section~\ref{sec_discussion}, 
and finally in Section~\ref{sec_conclusions} we present our conclusions.

\section{The Sample, Observations and Data Reduction}
\label{sec_sample}

 A brief description of the sample and the reasoning behind its selection is given
below -- this is described in greater detail in Paper~1.

The sample comprises 17 nearby `normal' (\ie\ systems without strong AGN)
spiral galaxies chosen to have one (or more) of the following
properties: ({\rm i}) They are X-ray bright, having an \Ein\ (0.2$-$4.0\,keV)
X-ray flux (Fabbiano \etal\ 1992) greater than $45\times10^{-13}$\,erg
cm$^{-2}$ s$^{-1}$, ({\rm ii}) they are infrared bright, having a 100\,$\mu$m
infrared flux (Soifer, Boehmer \& Neugebauer 1989) greater than 64\,Jy, and/or
({\rm iii}) they have a large value of `supernova flux' (\ie
$f_{SN}=L_{SN}/4\pi d^{2}$), where $L_{SN}$ is the estimated supernova
luminosity and $d$ is the distance) (see Read \& Ponman 1995) (note that many
of these galaxies are also found in the EUV sample of nearby spirals of Read
\& Ponman (1995)). 

Basic properties of the sample galaxies are shown in
Table\,\ref{table_sample}. For the sake of consistency, the galaxy distances
in Tully (1988), which are based on $H_{0} =
75$\,km~s$^{-1}$~Mpc$^{-1}$, and assume a Galaxy retardation of
300\,km~s$^{-1}$ from universal expansion by the mass of the Virgo cluster,
have been used. Axial ratios and major axis diameters are taken from the
Second Reference Catalog of Bright Galaxies (RCBG) (de Vaucouleurs, de
Vaucouleurs \& Corwin 1976). Optical (B) luminosities for all the galaxies in
the sample are taken from Tully (1988) and FIR luminosities are calculated
from IRAS 60 and 100\,$\mu$m fluxes (Soifer \etal\ 1989; Rice \etal\ 1988)
using the expression

\begin{displaymath}
L_{FIR} = 3.65\times10^{5}\left[2.58 S_{60\mu m} +
S_{100\mu m} \right] D^{2}  L_{\odot},
\end{displaymath}

(\eg Devereux \& Eales 1989), where $D$ is the distance (in Mpc) and $S_{60\mu
m}$ and $S_{100\mu m}$, the IRAS 60 and 100\,$\mu$m fluxes (in Janskys). An
asterisk by the galaxy name indicates a `starburst' galaxy, \ie\ one defined
as here with $L_{FIR}>0.38 L_{B}$. Also given in Table~\ref{table_sample} is
the FIR colour temperature $S_{60\mu m}/S_{100\mu m}$ (a good measure of the
activity), the galaxy type (\ie\ the mean numerical index of stage along the
Hubble sequence), from the RCBG, and, where applicable, the \Ein\ (0.2$-$4.0
keV) X-ray luminosity, calculated, for the distances tabulated, using the
fluxes given in Fabbiano \etal\ (1992), Trinchieri \etal\ (1988) and
Trinchieri \etal\ (1990). For convenience, the \Ros\ X-ray luminosity (as given 
in Table\,\ref{table_paper1}) is also given. The values of the Galactic hydrogen
column are taken from the neutral hydrogen radio results of Stark \etal\ (1992). 

\begin{table*}
\begin{center}
\begin{tabular}{lcccccccccc}        \hline
Galaxy & \multicolumn{4}{c}{Properties} & \multicolumn{4}{c}{Log
        Luminosity (erg s$^{-1}$)} &FIR temp. & Galactic \\ \cline{2-9}
  &Distance&Axial &Major & Type & $L_{B}$ & $L_{FIR}$ & 
   $L_{X}$    & $L_{X}$ & $S_{60\mu m}/S_{100\mu m}$ & Hydrogen \\
  &        & ratio& diam.&      &         &           & 
   (Einstein) & (Rosat) &                            & Column \\
  & (Mpc)& (D/d)&(\arcm)& & & & & & & $10^{20}$\,cm$^{-2}$ \\ \hline
N0055             & 1.3&5.01&32.4&9 &43.02&41.94&      & 38.82 &0.44 & 1.33 \\
N0247             & 2.1&2.69&20.0&7 &42.96&41.48&38.86 & 38.50 &0.29 & 1.33 \\
N0253*            & 3.0&3.39&25.1&5 &43.78&43.74&39.75 & 40.03 &0.50 & 1.23 \\
N0300             & 1.2&1.35&20.0&7 &42.52&41.43&      & 38.17 &0.31 & 2.97 \\
N0598, M33        & 0.7&1.58&61.6&6 &43.16&42.20&39.04 & 38.94 &0.33 & 4.56 \\
N0891*            & 9.4&4.79&13.5&3 &43.76&43.64&      & 40.36 &0.31 & 6.07 \\
N1291             & 8.6&1.15&10.5&0 &43.88&42.18&      & 40.07 &0.17 & 1.79 \\
N3034, M82*       & 5.2&2.45&11.2&I &43.54&44.25&40.88 & 40.85 &0.97 & 3.92 \\
N3079*            &20.4&4.47& 7.6&7 &44.18&44.14&40.47 & 40.62 &0.49 & 0.76 \\
N3628             & 7.7&4.07&14.8&3 &43.76&43.31&40.07 & 39.95 &0.49 & 1.77 \\
N4258, M106       & 6.8&2.29&18.2&4 &44.01&42.95&40.40 & 40.41 &0.27 & 1.30 \\
N4490*            & 7.8&1.91& 5.9&7 &43.59&43.26&      & 40.09 &0.52 & 1.61 \\
N4631*            & 6.9&4.57&15.1&7 &43.82&43.46&39.90 & 39.76 &0.40 & 1.18 \\
N5005*            &21.3&2.00& 5.4&4 &44.28&43.90&      & 40.72 &0.35 & 1.24 \\
N5055, M63        & 7.2&1.62&12.3&4 &43.83&43.28&      & 40.02 &0.25 & 1.24 \\
N5194, M51*       & 7.7&1.41&11.0&4 &44.07&43.66&40.39 & 40.57 &0.35 & 1.25 \\
N5457, M101       & 5.4&1.02&26.9&6 &43.95&43.30&39.83 & 39.76 &0.52 & 1.09 \\
\hline
\end{tabular}
\caption{Physical and multiwavelength 
properties of the survey sample (see text for details).}
\label{table_sample}
\end{center}
\end{table*}

Observations, data reduction and results for the individual 
systems are described in full in Paper~1. However, a brief 
summary of the reduction techniques is given here for convenience. 

All PSPC datasets (obtained from the UK \Ros\ Data Archive Centre at
the Department of Physics and Astronomy, Leicester University, UK) were 
reduced in exactly the same way using the Starlink {\em
ASTERIX} X-ray analysis system. 

In each case, the data were `cleaned' of high-background times, and the point
source search program PSS (Allan, Ponman \& Jeffries, ASTERIX User Note 4) 
was used
to search for point-like sources, the positions of which were then
cross-correlated with a variety of stellar and non-stellar catalogues. An
integrated (source + unresolved emission) spectrum was formed from a circular
area of diameter slightly greater than the optical diameter of each galaxy (so
as to include any diffuse emission above the plane in edge-on galaxies). 

A key aspect of the analysis performed in Paper~1, was the separation of the 
emission into diffuse and source components. 
In order to separate the diffuse from the source emission, data were removed at
the position of each PSS source, and the remaining data were then collapsed
into a spectrum and corrected for vignetting effects and exposure time.
To account for diffuse flux lost in the source-removal procedure, the
diffuse spectrum was renormalised using a `patched' image, where the holes
left after source removal were filled by bilinear interpolation. 
In Paper~1 we discussed tests of the reliability of these procedures for separating
pointlike and diffuse emission. As might be expected, we found a dependence
on the angular size of galaxies, and also on their activity.
The worst case is distant galaxies with strong diffuse emission. In this case
a significant fraction of the diffuse flux may be erroneously 
attributed to sources.
In the present sample, we estimate that at most three systems could
be affected at a level as large as 30\% of their diffuse flux. 
It should also be stressed that we have covered a large range in Hubble galaxy 
type, from galaxies with large elliptical-like bulges (Sa spirals) to galaxies 
with very small bulges and dominating spiral arms (Sd spirals). Furthermore, 
although the survey galaxies have been chosen as containing no dominating AGN, 
a modest, low-level range in AGN activity does exist across the survey.

Source
spectra were binned directly from the raw data, though in order to arrive at a
true {\em source} spectrum however, it was necessary to remove the flux from
both the local true background and the local diffuse emission, and this
required quite an intricate procedure (described in Paper~1). All source
spectra, integrated spectra and diffuse emission spectra were fitted with
standard spectral models. Power law and thermal bremsstrahlung models were
fitted to each of the point source spectra, whilst power law, Raymond \& Smith
(1977) hot plasma, and differential emission measure models (Raymond and Smith
models integrated over a range of temperatures) were fitted to the diffuse and
integrated spectra. Minimum chi-squared fitting was used for the diffuse and
integrated spectra, and so a goodness of fit could be obtained. For the point
source spectra, which generally contained far fewer counts, Gaussian statistics
cannot be assumed, so a maximum likelihood criterion was used.

Assuming the diffuse emission to be hot gas, basic physical properties of
this gas could be inferred on the basis of assumptions about the
geometry of the emission. In the {\em bubble} model, where the gas is
assumed to be contained in a spherical bubble of radius $r$, the fitted
emission measure $\eta n^{2}_{e} V$ (where $\eta$ is the `filling factor' - the
fraction of the total volume $V$ which is occupied by the emitting gas) can be
used to infer the mean electron density $n_{e}$, and hence the total mass
$M_{\mbox{\small gas}}$ ($\propto n_{e} V$), thermal energy 
$E_{\mbox{\small th}}$ ($\propto n_{e} V T$, where $T$ is the temperature), and cooling time 
$t_{\mbox{\small cool}}$ ($\propto E_{\mbox{\small th}} L_{X}^{-1}$) of the
hot gas.

Table\,\ref{table_paper1} presents a summary of the results obtained in 
Paper~1. After the galaxy name (asterisks again indicating the `starburst'
galaxies), is given 
the integrated (i.e.\,source plus diffuse) 
(0.1$-$2.0\,keV) X-ray luminosity (also given in Table\,\ref{table_sample}). 
The next two columns give 
the number of sources detected within the galaxy and the integrated source 
X-ray luminosity. The remaining eight columns deal with the diffuse emission. 
We list the X-ray luminosity of the diffuse emission, the `diffuse fraction' 
(the ratio of $L_{X}^{\mbox{\small diff}}$ to $L_{X}^{\mbox{\small diff}}$ plus 
$L_{X}^{\mbox {\small src}}$), and the fitted temperature of the diffuse emission. The
last five columns deal with the `bubble' model gas parameters: the
radius $r$ of the assumed bubble is taken as the average radius of the lowest
contour level seen in each diffuse emission image in Paper~1; this is followed
by the mean electron density, total mass, thermal energy and cooling time of 
the hot gas.

\begin{table*}
\begin{center}
\begin{tabular}{lccccccccccc}    \hline

Galaxy & Integrated  & \multicolumn{2}{c}{Source emission} & 
         \multicolumn{8}{c}{Diffuse emission} \\ \hline 

       & Log $L_{X}$ & $N_{\mbox{\small src}}$ & Log $L_{X}^{\mbox{\small src}}$   & 
         Log $L_{X}^{\mbox{\small diff}}$ & $f_{\mbox{\small diff}}$ & $T_{\mbox{\small diff}}$ & $r_{\mbox{\small diff}}$ & 
         $n_{e}$ & $M_{\mbox{\small gas}}$ & $E_{\mbox{\small th}}$ & 
         $t_{\mbox{\small cool}}$ \\

       & (0.1$-$2.0\,keV) &             & (0.1$-$2.0\,keV)           & 
         (0.1$-$2.0\,keV)          &                 & (keV)         & (kpc) & 
         (cm$^{-3}$)& ($M_{\odot}$)       &  (erg)                  & 
         (Myr)                     \\

       & (erg s$^{-1}$)  & & (erg s$^{-1}$)  & 
         (erg s$^{-1}$)            &                 & & & 
         ($\times1/\sqrt{\eta}$) & ($\times\sqrt{\eta}$) & ($\times\sqrt{\eta}$) & 
         ($\times\sqrt{\eta}$) \\ \hline

N0055 & 38.82 &  8 & 38.71 & 38.16 & 0.22 &      &      &        &                   &                    &      \\
N0247 & 38.50 &  5 & 38.35 & 37.97 & 0.29 & 0.16 & 2.7  & 0.0030 & $3.1\times10^{6}$ & $2.8\times10^{54}$ & 530  \\
N0253*& 40.03 & 15 & 39.45 & 39.90 & 0.74 & 0.47 & 9.2  & 0.0062 & $2.5\times10^{8}$ & $6.6\times10^{56}$ & 2620 \\
N0300 & 38.17 & 15 & 38.13 & 37.19 & 0.10 &      &      &        &                   &                    &      \\
N0598 & 38.94 & 36 & 38.84 & 38.27 & 0.21 & 0.43 & 1.4  & 0.0228 & $3.0\times10^{6}$ & $7.2\times10^{54}$ & 1230 \\
N0891*& 40.36 &  2 & 39.68 & 40.26 & 0.79 & 0.11 & 9.0  & 0.0022 & $8.2\times10^{7}$ & $5.0\times10^{55}$ & 90   \\
N1291 & 40.07 &  4 & 39.72 & 39.81 & 0.55 & 0.55 & 8.6  & 0.0063 & $2.1\times10^{8}$ & $6.4\times10^{56}$ & 3130 \\
N3034*& 40.85 &  2 & 40.35 & 40.69 & 0.69 & 0.72 & 7.6  & 0.0234 & $5.4\times10^{8}$ & $2.2\times10^{57}$ & 1380 \\
N3079*& 40.62 &  1 & 40.12 & 40.46 & 0.69 & 0.52 & 14.8 & 0.0059 & $1.0\times10^{9}$ & $2.9\times10^{57}$ & 3160 \\
N3628 & 39.95 &  7 & 39.82 & 39.37 & 0.26 & 0.36 & 9.0  & 0.0035 & $1.3\times10^{8}$ & $2.7\times10^{56}$ & 3590 \\
N4258 & 40.41 &  3 & 39.70 & 40.32 & 0.81 & 0.35 & 7.4  & 0.0098 & $2.1\times10^{8}$ & $4.0\times10^{56}$ & 610  \\
N4490*& 40.09 &  4 & 39.74 & 39.83 & 0.55 &      &      &        &                   &                    &      \\
N4631*& 39.76 &  7 & 39.35 & 39.56 & 0.62 & 0.23 & 8.4  & 0.0034 & $1.0\times10^{8}$ & $1.3\times10^{56}$ & 930  \\
N5005*& 40.72 &  3 & 40.50 & 40.32 & 0.40 & 0.42 & 9.9  & 0.0080 & $4.1\times10^{8}$ & $9.4\times10^{56}$ & 1430 \\
N5055 & 40.02 &  6 & 39.64 & 39.78 & 0.58 & 0.58 & 6.7  & 0.0103 & $1.6\times10^{8}$ & $5.2\times10^{56}$ & 2730 \\
N5194*& 40.57 &  9 & 40.20 & 40.33 & 0.57 & 0.51 & 11.2 & 0.0084 & $6.2\times10^{8}$ & $1.7\times10^{57}$ & 2580 \\
N5457 & 39.76 & 13 & 39.44 & 39.48 & 0.52 & 0.26 & 17.3 & 0.0011 & $2.9\times10^{8}$ & $4.1\times10^{56}$ & 4250 \\

\hline
\end{tabular}
\caption{Physical properties derived from the X-ray analysis, summarised
from Paper~1, including integrated (total) and point source and diffuse
luminosities, number of X-ray sources detected, and a variety of inferred
properties of the hot gas (see text for details).}
\label{table_paper1}
\end{center}
\end{table*}

\section{Results}
\label{sec_results}

Correlation tests were performed between almost all of possible pairs of
the parameters listed in tables~\ref{table_sample} and \ref{table_paper1}, and
for a variety of galaxy selections. The most interesting 
results of regression and Spearman rank correlation analyses
performed on the data are given in
Table~\ref{table_slopes} as follows: The Y and X parameters (cols.\,1 \& 2),
the galaxy selection criteria used (whether all the galaxies, just the
starbursts [S], just the normal galaxies [N], or any individual omissions
[\eg\ no M82]) (col.\,3), and the values and associated 1$\sigma$ errors for
the regression gradient $m$ (col.\,4) and constant $c$ (col.\,5). Here the
assumed regression is of the form $Y=mX+c$. Also given (col.\,6), is the 
Spearman rank-order correlation coefficient $r_{S}$ (lying between -1 and +1), 
and (col.\,7) the associated significance of $r_{S}$, $T_{S}$. Here, $T_{S}$ is 
distributed approximately as Student's distribution with $N-2$ degrees of freedom. 
Only fit results for which the regression software produced a stable
solution are listed in Table~\ref{table_slopes}.
For example, for the $\log{L_{X}}:\log{L_{B}}$ relation, no fit for 
the starburst subsample could be derived, though a stable solution
was obtained when M82 was omitted from the sample, and this is listed.
Further results, for instance Spearman rank correlation coefficients for
pairs of parameters where no regression fits were found, are given in the following text.

\begin{table*}
\begin{center}
\begin{tabular}{llccccc}  \hline
Y & X & Sel.[$N$] & m & c & $r_{S}$ & $T_{S}$ \\ \hline

$\log{L_{X}}$ & $\log{L_{FIR}}$ & All [17] & 0.80$\pm$0.10 & 5.38$\pm$0.37 & 0.74 & 4.2 \\
                    & & S [8] & 0.90$\pm$0.30 & 1.08$\pm$0.26 & 0.80 & 3.2 \\
                    & & N [9] & 0.87$\pm$0.22 & 2.48$\pm$0.47 & 0.65 & 2.2 \\

$\log{L_{X}^{\mbox{\small diff}}}$ & $\log{L_{FIR}}$   & All [17] & 1.01$\pm$0.13 & -3.88$\pm$0.48 & 0.80 & 5.2 \\
                                    & & S [8] & 0.93$\pm$0.26 & -0.67$\pm$0.22 & 0.86 & 4.2 \\
                                    & & N [9] & 1.10$\pm$0.30 & -7.92$\pm$0.66 & 0.63 & 2.2 \\

$\log{L_{X}^{\mbox{\small src}}}$ & $\log{L_{FIR}}$   & All [17] & 0.66$\pm$0.09 & 11.20$\pm$0.33 & 0.74 & 4.2 \\
                                    & & S [8] & 0.83$\pm$0.40 & 3.54$\pm$0.35 & 0.64 & 2.0 \\
                                    & & N [9] & 0.73$\pm$0.16 & 8.33$\pm$0.35 & 0.78 & 3.3 \\

$\log{L_{X}/L_{FIR}}$ & $S_{60\mu m}/S_{100\mu m}$ & All [17] & -1.14$\pm$0.49 & -2.74$\pm$0.35 & -0.56 & -2.6 \\
                                               & & N [9] & -2.84$\pm$0.97 & -2.08$\pm$0.32 & -0.75 & -3.0 \\

$\log{L_{X}}$ & $\log{L_{B}}$  & All [17] & 1.48$\pm$0.20 & -24.87$\pm$0.38 & 0.60 & 2.9 \\
           & & S, no M82 [7] & 1.11$\pm$0.39 & -8.49$\pm$0.24 & 0.70 & 2.2 \\
                   & & N [9] & 1.44$\pm$0.12 & -23.26$\pm$0.19 & 0.90 & 5.6 \\

$\log{L_{X}^{\mbox{\small diff}}}$ & $\log{L_{B}}$ & All [17] & 1.89$\pm$0.25 & -43.06$\pm$0.48 & 0.49 & 2.2 \\
                                & & S, no M82 [7] & 0.92$\pm$0.42 & -0.33$\pm$0.26 & 0.66 & 2.0  \\
                                & & N [9] & 1.91$\pm$0.14 & -43.88$\pm$0.22 & 0.95 & 8.2 \\

$\log{L_{X}^{\mbox{\small src}}}$ & $\log{L_{B}}$     & All [17] & 1.23$\pm$0.18 &-14.05$\pm$0.34 & 0.64 & 3.2 \\
           & & S, no M82 [7] & 1.37$\pm$0.43 & -20.34$\pm$0.26 & 0.63 & 1.8 \\
                   & & N [9] & 1.15$\pm$0.13 & -10.72$\pm$0.19 & 0.76 & 3.1 \\

$\log{L_{FIR}}$ & $\log{L_{B}}$  & All [17] & 1.48$\pm$0.30 & -21.63$\pm$0.56 & 0.51 & 2.3 \\
           & & S, no M82 [7] & 0.92$\pm$0.29 &  3.13$\pm$0.18 & 0.80 & 3.0 \\
                   & & N [9] & 1.22$\pm$0.28 & -10.63$\pm$0.42 & 0.71 & 2.7 \\

$\log{L_{X}^{\mbox{\small diff}}/L_{B}}$ & $\log{L_{B}}$ & All [17] & 0.89$\pm$0.25 & -43.09$\pm$0.48 & 0.50 & 2.2 \\
                                & & N [9] & 0.91$\pm$0.14 & -43.88$\pm$0.22 & 0.87 & 4.7 \\

$\log{L_{X}^{\mbox{\small diff}}/L_{B}}$ & $\log{L_{FIR}/L_{B}}$ & All [17] & 0.75$\pm$0.19 & -3.70$\pm$0.45 & 0.71 & 3.9 \\
                                & & S [8] & 0.96$\pm$0.23 & -3.59$\pm$0.23 & 0.58 & 1.8 \\

$\log{L_{X}^{\mbox{\small src}}/L_{B}}$ & $\log{L_{FIR}/L_{B}}$ & All [17] & 0.22$\pm$0.05 & -4.28$\pm$0.23 & 0.55 & 2.6 \\

$\log{r_{\mbox{\small diff}}}$ & $\log{L_{B}}$ & [14] & 0.65$\pm$0.12 & -27.51$\pm$0.16 & 0.66 & 3.0 \\
                                & & S [7]                 & 0.28$\pm$0.10 & -11.08$\pm$0.07 & 0.80 & 3.0 \\
                                & & N [7]                 & 0.76$\pm$0.20 & -32.34$\pm$0.20 & 0.63 & 1.8 \\



$f_{\mbox{\small diff}}$ & $\log{L_{B}}$ & All [17] & 0.32$\pm$0.09 & -13.41$\pm$0.16 & 0.43 & 1.8 \\
                            & & N [9] & 0.36$\pm$0.09 & -15.14$\pm$0.13 & 0.82 & 3.8 \\





$\log{L_{X}^{\mbox{\small diff}}}$ & $\log{T_{\mbox{\small diff}}}$ & [14] & 
2.5$\pm$1.6 & 40.84$\pm$0.69 & 0.48 & 1.9 \\


$\log{M_{\mbox{\small gas}}}$ & $\log{L_{FIR}}$ & [14] & 0.77$\pm$0.16 & -25.24$\pm$0.47 & 0.70 & 3.4 \\
                                                 & & S [7] & 1.05$\pm$0.43 & -37.34$\pm$0.30 & 0.70 & 2.2 \\
                                                 & & N [7] & 0.94$\pm$0.35 & -32.44$\pm$0.63 & 0.40 & 1.0 \\


 



$\log{L_{X}/L_{B}}$ & Type & no M82 [16] & -0.075$\pm$0.035 & -3.48$\pm$0.31 & -0.44 & -1.8 \\
                       & & N [9] & -0.083$\pm$0.028 & -3.63$\pm$0.21 & -0.71 & -2.7  \\






\hline
\end{tabular}
\caption{Main results of regression analyses (assuming slopes of the form $Y
= mX + c$, unweighted on $Y$ and $X$) performed on the galaxy sample. The Table
lists the $Y$ and $X$
parameters, the galaxy selection criteria used, and the values and associated
1$\sigma$ errors for the regression gradient $m$ and the intercept $c$. Also
given is the Spearman rank-order correlation coefficient $r_{S}$ and its
associated significance $T_{S}$ (see text for details).
}
\label{table_slopes}
\end{center}
\end{table*}

From amongst the large number of available relations, we 
concentrate below on those which are available here for the first time, as a result
of our analysis: the individual source and diffuse contributions to the X-ray 
emission, and properties derived from these. We examine the 
variation of these components with galaxy mass and activity, discuss the relation
to emission in other wavebands, and finally investigate some of the derived
gas parameters, and the variation in X-ray properties with Hubble galaxy type.

Scatter plots follow a standard format:
individual galaxy names are given, centred on their particular positions in
the relevant parameter space (brackets indicate that the parameter position of
the galaxy lies closer to the nearest non-bracketed galaxy than is reasonable
to plot). Starburst galaxies are shown in a bolder font and underlined. 
Where relevant, the regression fits from Table~\ref{table_slopes}, for the
full sample (bold line), and for the (S)tarburst and (N)ormal subsamples
(dashed lines) are also plotted.

\subsection{Mass and activity}
\label{sec_massact}

Before proceeding further, we need to clarify how we define
the `mass' and `activity' of our galaxies.
We take the optical luminosity, $L_{B}$, as our measure of mass.
This is not ideal, since the blue luminosity to mass ratio is quite sensitive
to the age of a stellar population. Luminosity in the near infrared would be
a more robust indicator of stellar mass, but is not available
for many of the galaxies in our sample.
Several studies over the years (\eg\ Devereux \& Young 1991; Joseph \etal\ 1984) 
have linked the
far-infrared luminosity of a galaxy to the star-formation rate (SFR), and
hence (for a given initial mass function) to the supernova rate.
We therefore take these rates to be proportional to $L_{FIR}$.
Therefore, we can also consider the `activity',
\ie\ the `SFR per unit mass', to be measured by $L_{FIR}/L_{B}$. 

Given the parameters at our disposal, we could consider other indicators
of mass and activity. For example, the far-infrared colour temperature
$S_{60\mu m}/S_{100\mu m}$ (see \eg\ Telesco \etal\ 1988) is an attractive
activity indicator, except that it has rather large statistical errors
for many of our galaxies. An alternative possible measure for the
mass might be galaxy volume, defined as (diameter$^2\times$axis-ratio)$^{1.5}$.
Comparing these alternative measures, it is encouraging to note that they
are well correlated, as can be seen in Fig.\,1 (top and middle) -- $L_{B}$ 
is strongly correlated to galaxy volume ($T_{S}=3.2$), and $L_{FIR}/L_{B}$ 
with the far-infrared colour temperature ($T_{S}=2.5$). This 
strengthens our confidence that the chosen parameters are good
tracers of mass and activity. 

Since we wish to separate the effects of mass and activity on galaxy
properties, it will help a great deal if mass and activity are independent
variables for our galaxy sample. If, for example, large galaxies in the sample
tended to have higher activity than small ones, it would be difficult to
disentangle the effects of the two variables. Fortunately, there is
no strong correlation between the mass tracer and the activity tracer
(see Fig.\,1 (bottom)). Our sample does appear to be
lacking in high-$L_{FIR}$-low-$L_{B}$ systems, 
\ie\ high-activity dwarfs. This is very likely due to them being rare, and hence 
distant, and thus not meeting the flux limits set for the sample. Despite this,
the Spearman rank significance, $T_{S}$, for any correlation between
our mass and activity parameters lies well below 1 not only for the total sample, 
but also for the normal and starburst subsamples. This makes the analysis of 
the following results a good deal easier. 

\begin{figure}
\unitlength1.0cm
    \psfig{figure=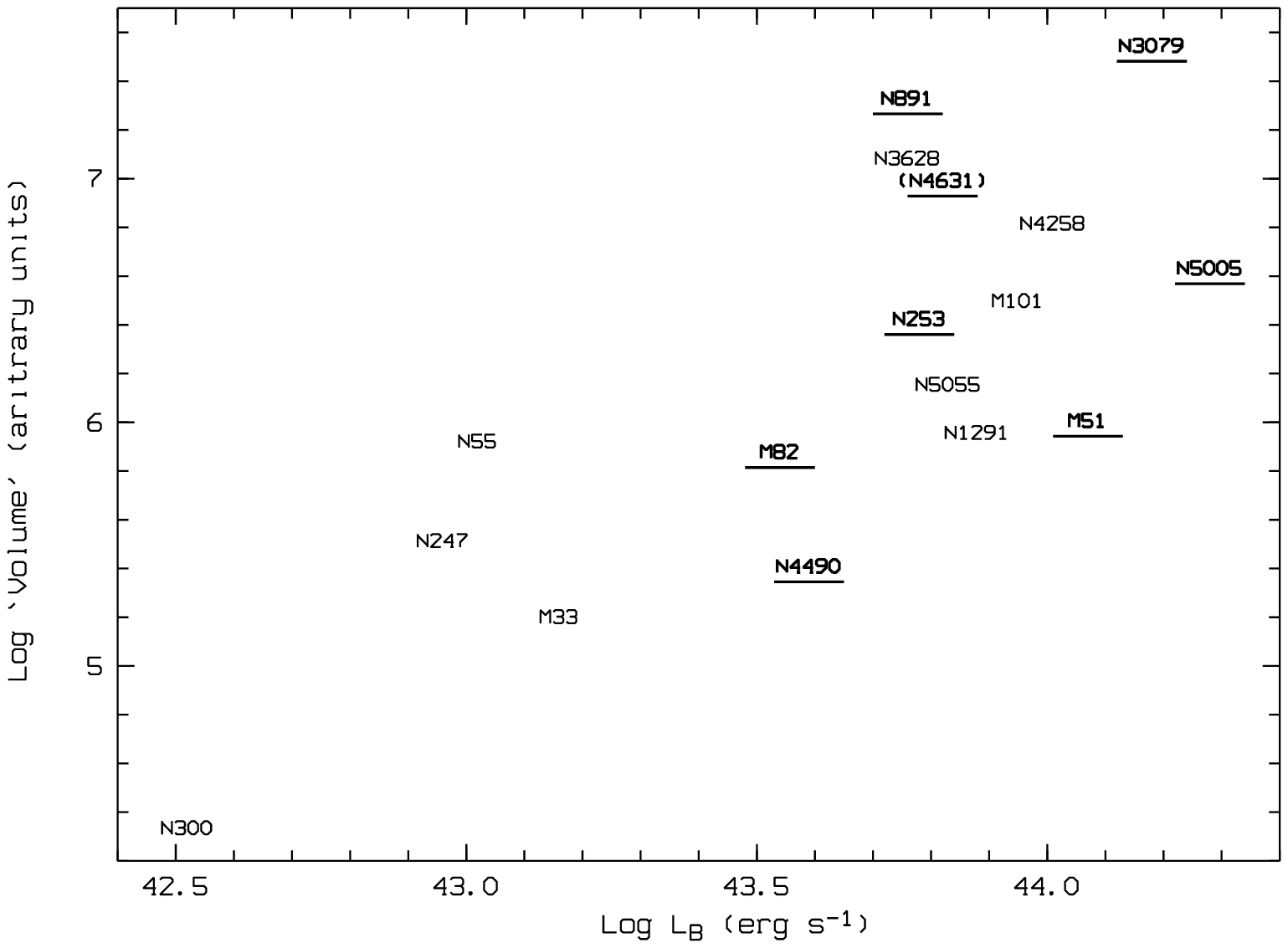,width=9.0cm,clip=}
    \psfig{figure=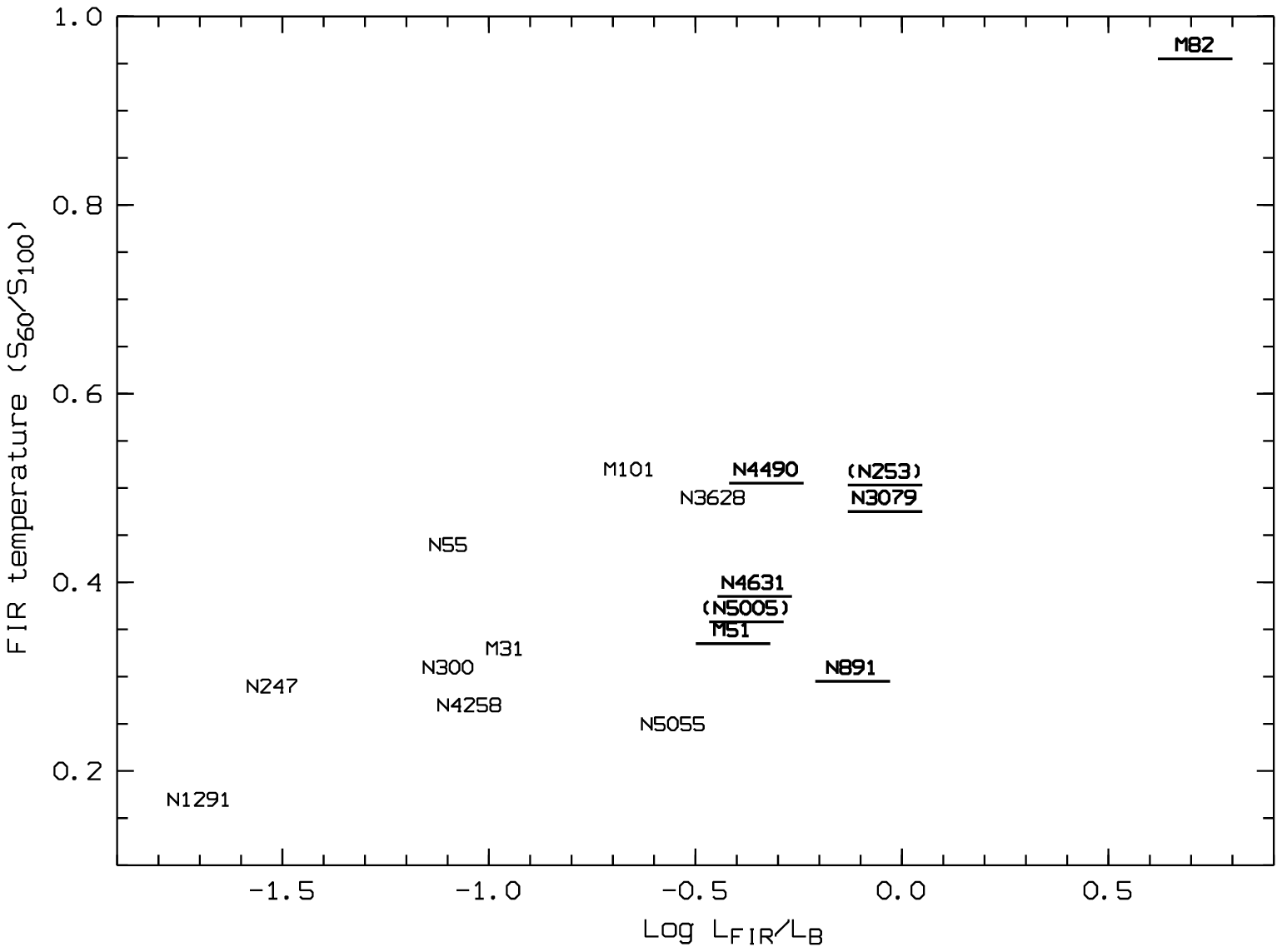,width=9.0cm,clip=}
    \psfig{figure=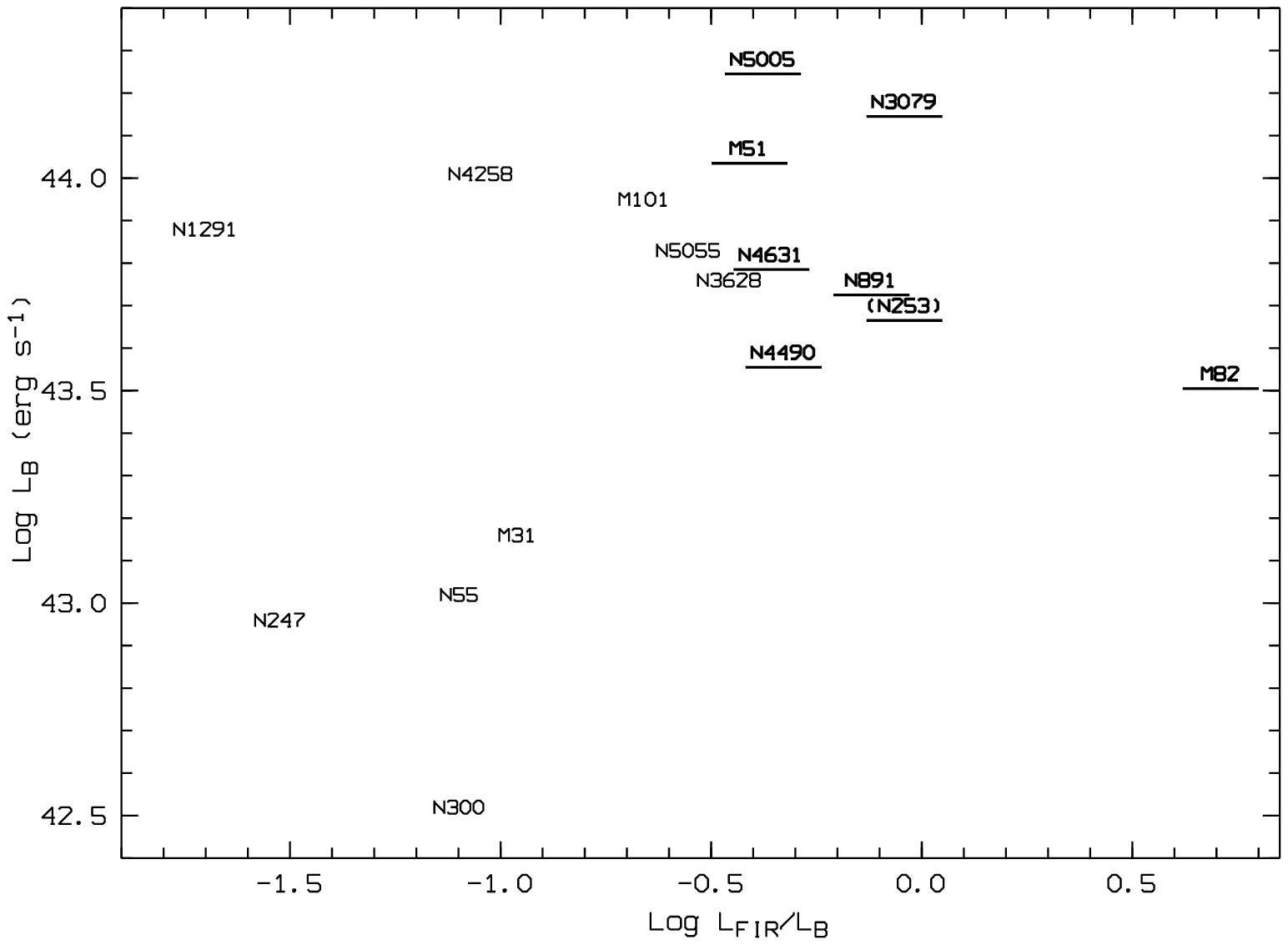,width=9.0cm,clip=}
\hfill \parbox[b]{9.0cm}
\caption{
Relationships between (top) the two suggested `mass' parameters, 
$L_{B}$ and `volume', (middle) the two suggested `activity' 
parameters, $L_{FIR}/L_{B}$ and the far-infrared colour temperature, and 
(bottom) the adopted mass and activity parameters, $L_{B}$ and $L_{FIR}/L_{B}$. 
}
\end{figure}
\label{fig_massact}

\subsection{Variations in X-ray properties with mass and activity}
\label{sec_X-massact}

First, we compared the different components of the X-ray emission with 
the galaxy mass for the total sample, and for the normal and starburst 
subsamples. Fig.\,2 shows the variation in the diffuse X-ray emission 
fraction ($f_{\mbox{\small diff}}$) with the `mass' ($L_{B}$). As can be seen
in the figure (and in Table \ref{table_slopes}), a gradual increase in 
the diffuse X-ray emission fraction with mass is observed, i.e. high mass 
galaxies have a smaller fraction of their X-ray emission in discrete sources.
This relationship is driven by the normal subsample, and is not seen
in the starbursts. 

\begin{figure}
\unitlength1.0cm
    \psfig{figure=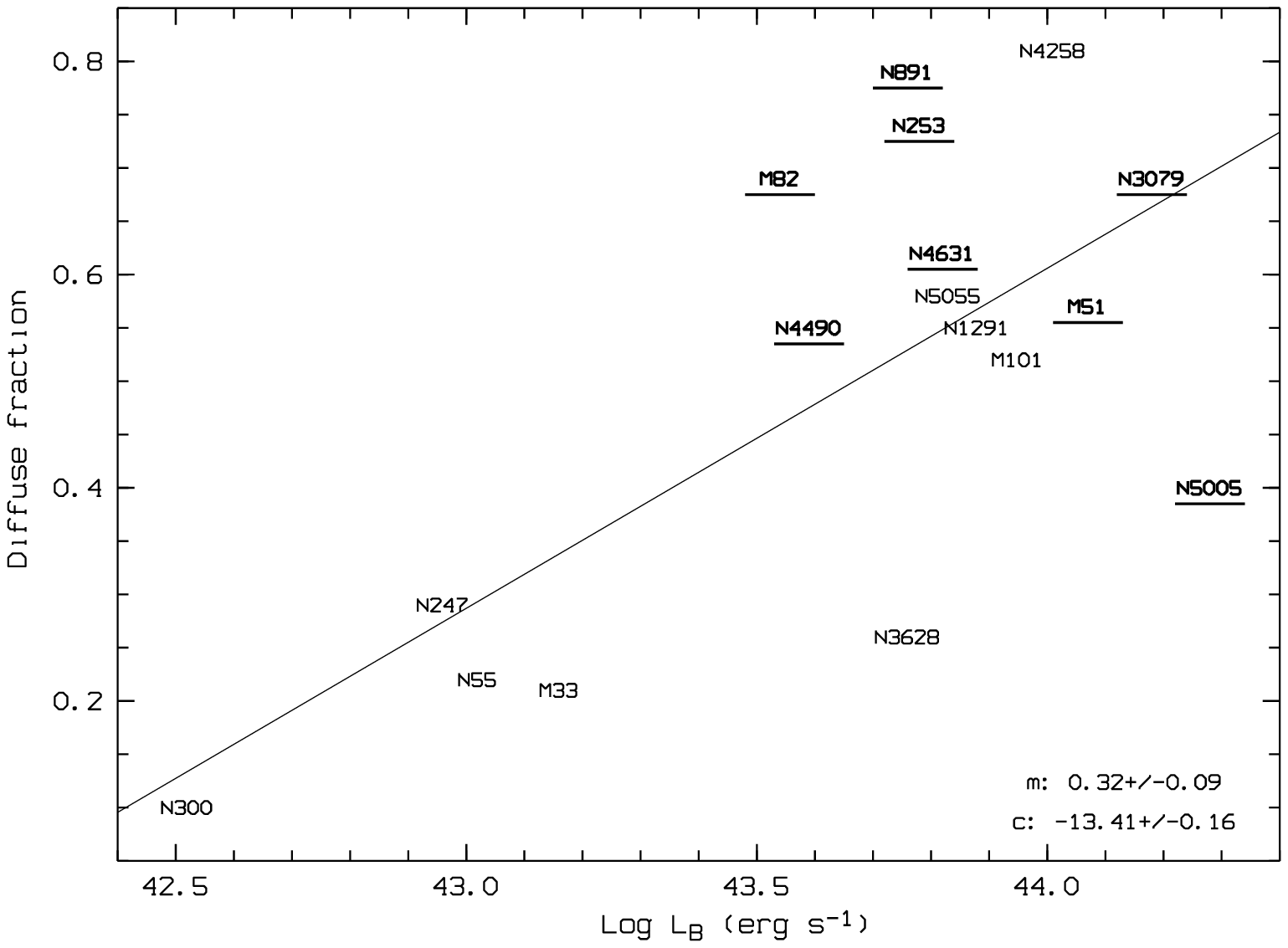,width=9.0cm,clip=}
\hfill \parbox[b]{9.0cm}
\caption{
The variation in the diffuse X-ray emission
fraction ($f_{\mbox{\small diff}}$) with galaxy `mass' ($L_{B}$).
}
\label{difffrac_lb}
\end{figure}

Comparing source and diffuse X-ray luminosity {\em per unit 
galaxy mass} (\ie $L_{X}^{\mbox{\small src}}/L_{B}$ and 
$L_{X}^{\mbox{\small diff}}/L_{B}$) with galaxy mass, we find no
significant trend in {\it source} luminosity (per 
unit galaxy mass). However,
the production of {\it diffuse} X-ray emission per unit galaxy mass is 
seen to correlate strongly with galaxy mass (see Table \ref{table_slopes}) 
for the normal galaxy subsample ($T_{S}=4.5$).

The same X-ray parameters were compared with the activity tracer, 
$L_{FIR}/L_{B}$. Diffuse fraction ($f_{\mbox{\small diff}}$) 
is seen to rise with activity within the starburst
sample ($T_{S}=2.4)$, but shows no significant trend within the normal galaxy sample.
Diffuse X-ray luminosity per unit galaxy mass ($L_{X}^{\mbox{\small diff}}/L_{B}$),
Fig.\,3 (top), shows a trend with activity across the whole galaxy sample. 
Starbursts have (by definition) high activity, and also higher diffuse X-ray emission
than normal galaxies.

\begin{figure}
\unitlength1.0cm
    \psfig{figure=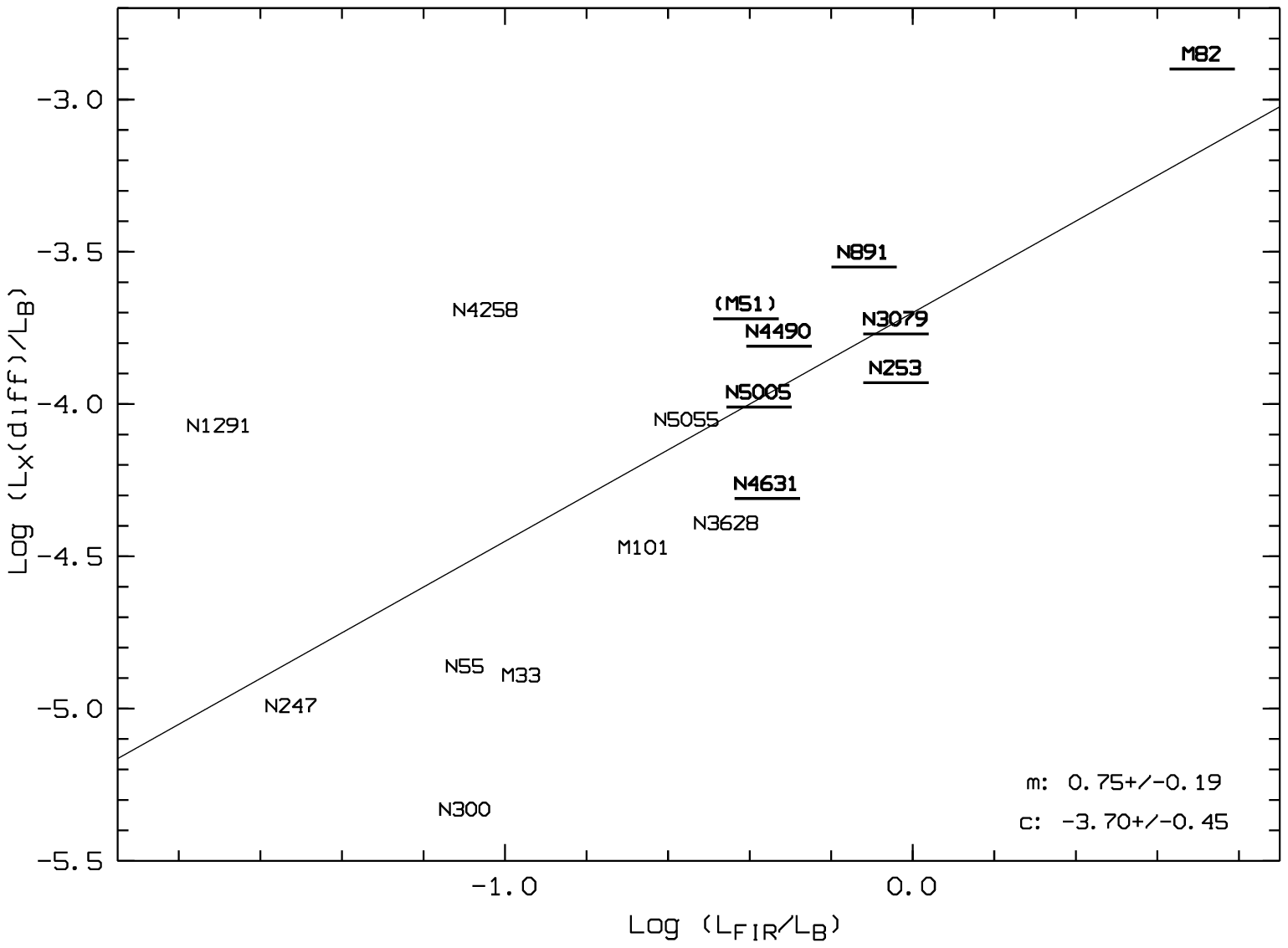,width=9.0cm,clip=}
    \psfig{figure=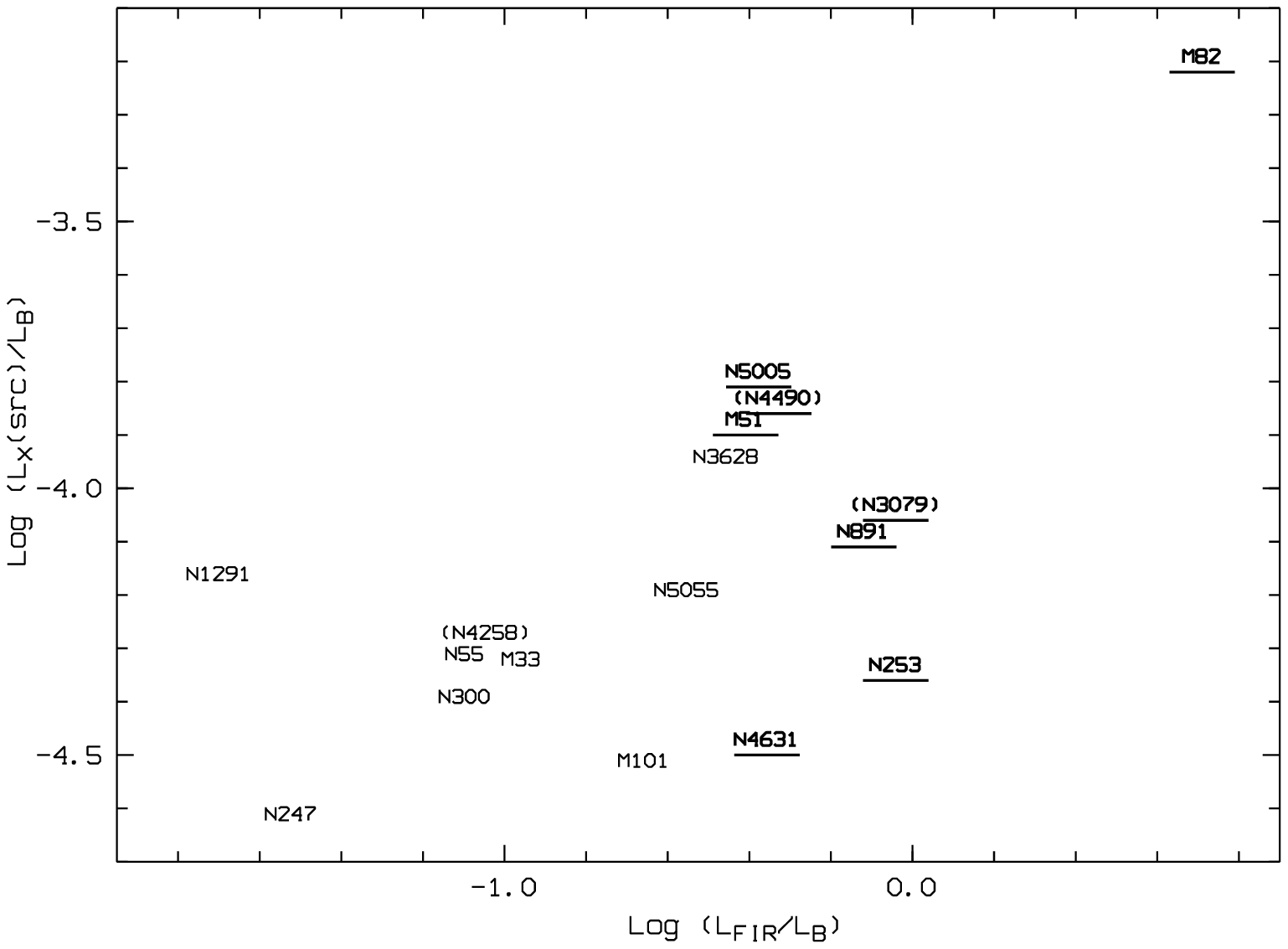,width=9.0cm,clip=}
\hfill \parbox[b]{9.0cm}
\caption{
The variation of (top) the diffuse X-ray luminosity per unit galaxy mass
($L_{X}^{\mbox{\small diff}}/L_{B}$) and (bottom) the source X-ray luminosity per
unit galaxy mass ($L_{X}^{\mbox{\small src}}/L_{B}$) with the activity
($L_{FIR}/L_{B}$) for the galaxy sample.

}
\label{lxdifflb_logfiropt}
\end{figure}

The {\em source} 
X-ray luminosity per unit galaxy mass (see Fig.\,3 (bottom)) does
show a general rise with activity from
normal to starburst galaxies within the total sample ($T_{S}=2.6$), though the
trend is not significant within the individual subsamples ($T_{S}<1$ in both cases).

The broad picture which emerges from the above is that normal galaxies appear to 
be very much more influenced by their mass than by their activity. 
For starburst galaxies, on the other hand, it is activity that 
is the dominating factor.

We will discuss the properties of the diffuse X-ray emission in detail later, but
comment here on some of the more noteworthy effects of galaxy mass and activity on 
the properties of the gas. The derived temperatures of the
diffuse X-ray emission shows no really significant trends with either 
mass or activity. Only a low-significance positive 
correlation in diffuse emission temperature with activity is observed, 
and then only in the starburst subsample ($T_{S}=1.3$). 
The radial extent of the diffuse emission does show significant 
positive correlations with galaxy mass. Fig.\,4 shows the relationship
between the log of the radial extent of the diffuse emission ($r$; see 
Table~\ref{table_paper1}) and the galaxy mass tracer $L_{B}$. 
As can be seen from the figure, and from 
Table~\ref{table_slopes}, a positive trend is seen, both in the total sample, 
and in the individual starburst and normal subsamples. Larger galaxies  
posess larger diffuse emission features. 

\begin{figure}
\unitlength1.0cm
    \psfig{figure=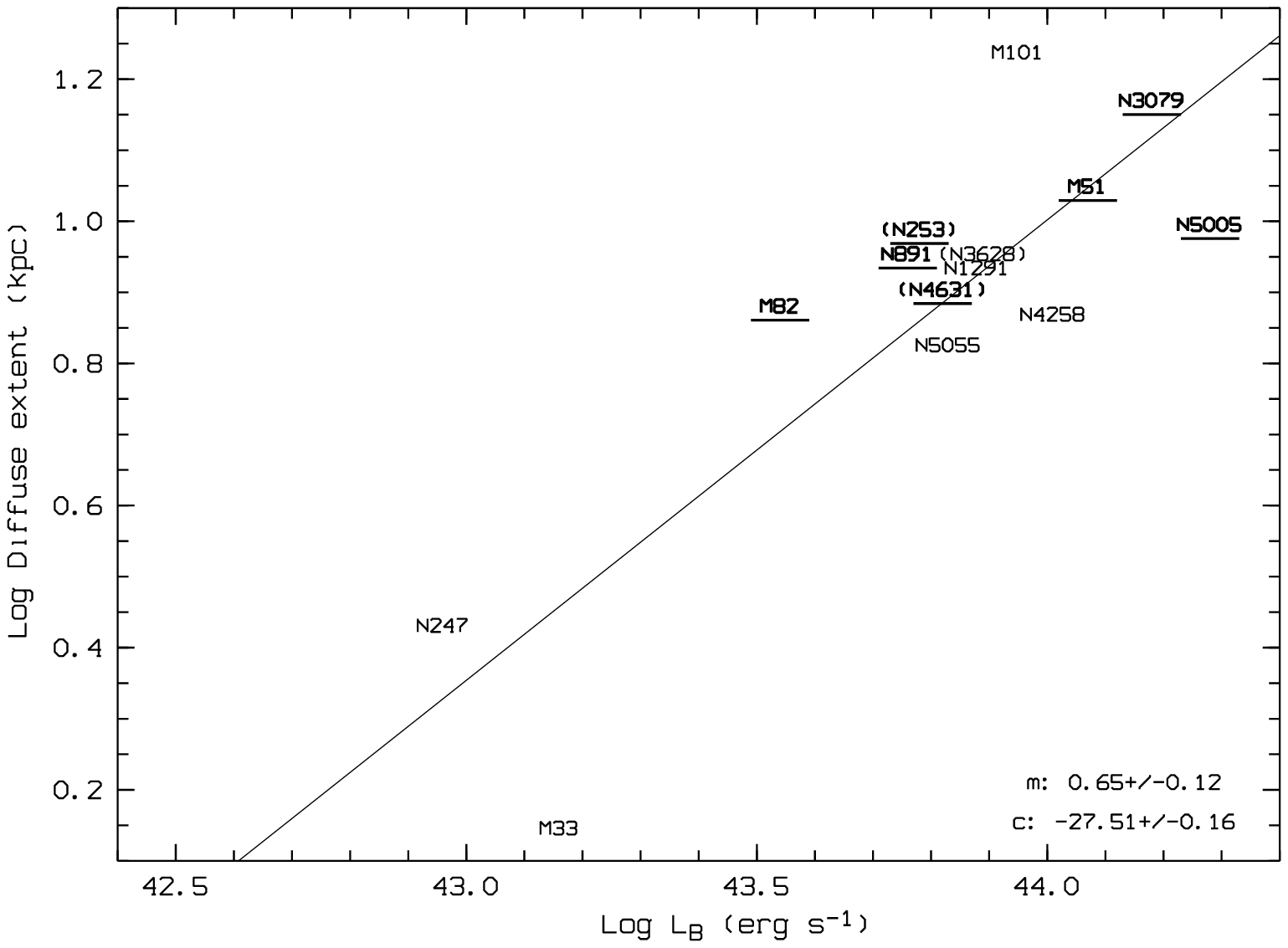,width=9.0cm,clip=}
\hfill \parbox[b]{9.0cm}
\caption{
The relationship
between the log of the radial extent of the diffuse emission ($r$; see 
Table~\ref{table_paper1})
and the galaxy mass tracer $L_{B}$ for the total sample.
}
\label{radius_lb}
\end{figure}

Surprisingly, only a low-significance positive correlation is seen between the radial
extent of the diffuse emission and galaxy activity. For the total sample, only a 
Spearman rank coefficient of 0.42 (corresponding to $T_{S}$: 1.6) is obtained. Nothing
significant is seen within the individual normal and starburst galaxy subsamples.

\subsection{X-ray$-$Far-Infrared luminosity relationships}
\label{sec_X-FIR}

Since $L_{FIR}$ is a measure of star formation rate, a positive relation
with X-ray emission is to be expected. Fig.~5 shows the
relationships between the various components of the X-ray emission and the
far-infrared luminosity. \Ros\ (0.1$-$2.0\,keV) luminosities of the total
(top), the diffuse (middle) and the source (bottom) emission components are
shown plotted against the far-infrared luminosity for the survey sample. 

\begin{figure}
\unitlength1.0cm
    \psfig{figure=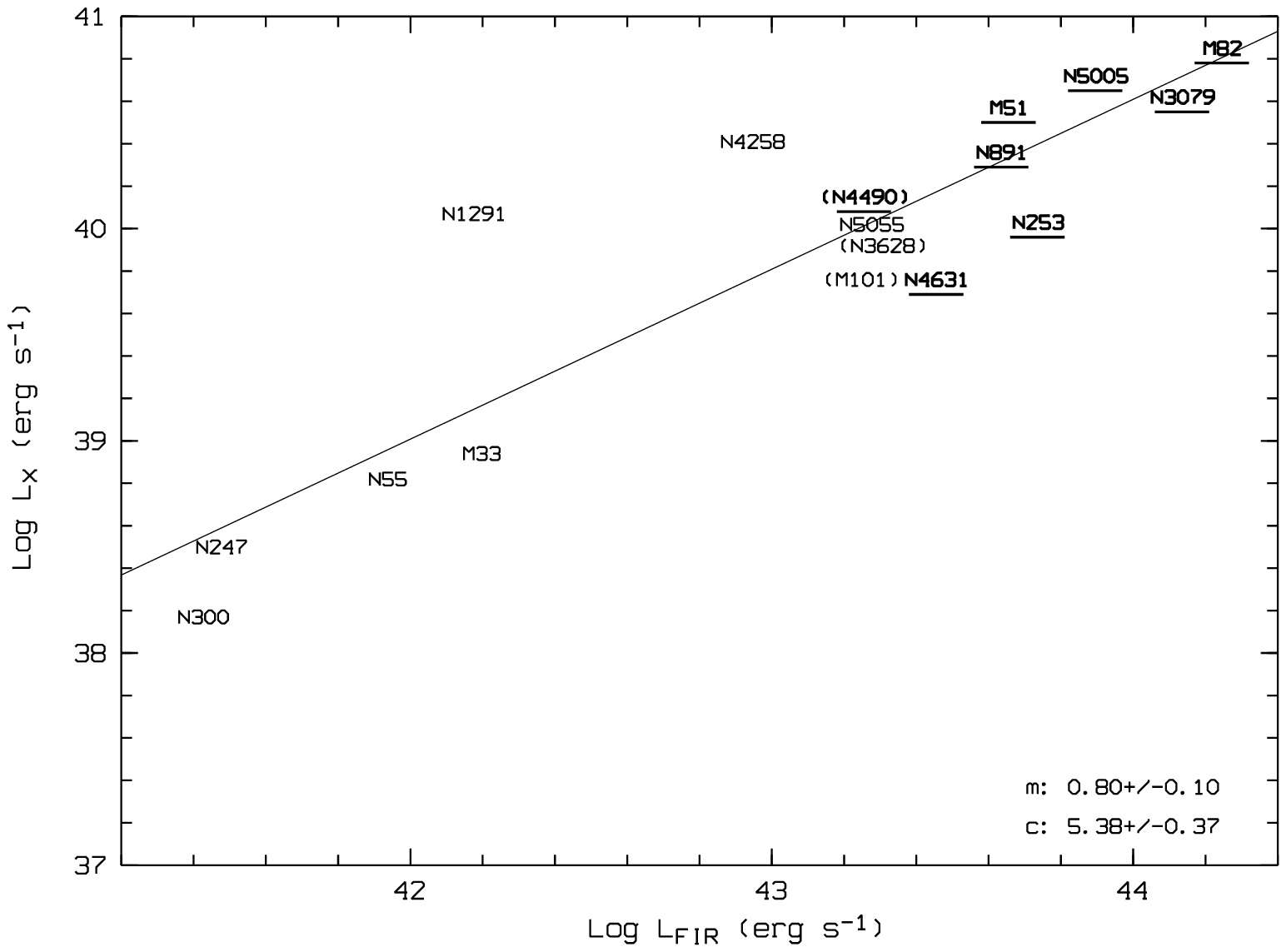,width=9.0cm,clip=}
    \psfig{figure=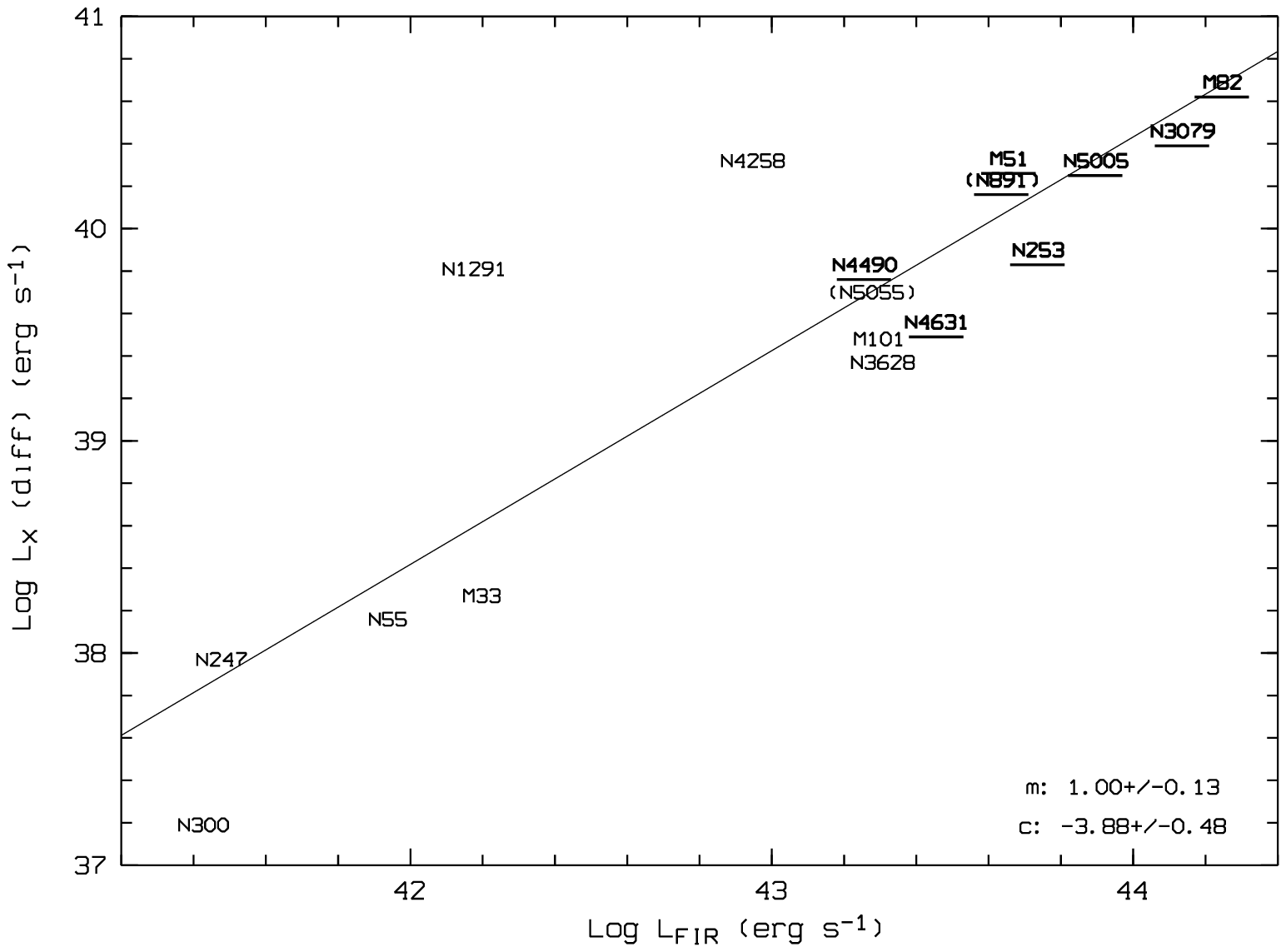,width=9.0cm,clip=}
    \psfig{figure=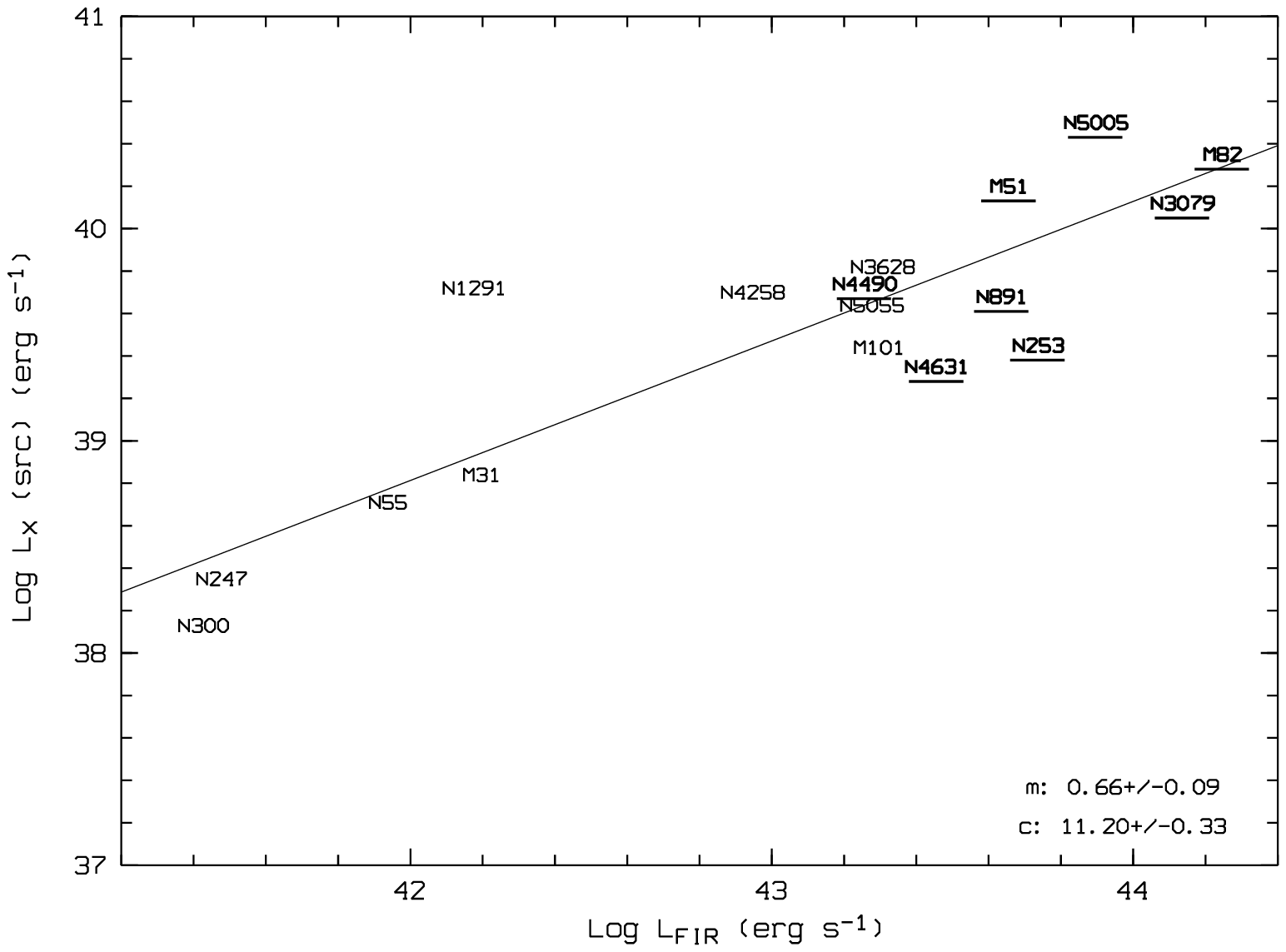,width=9.0cm,clip=}
\hfill \parbox[b]{9.0cm}
\caption{
\Ros\ (0.1$-$2.0\,keV) total (top), diffuse (middle) and source (bottom) 
X-ray luminosity versus far-infrared luminosity
for the survey sample. Starburst galaxies are marked in a bold font and 
underlined.
Regression fit lines to the total sample (bold line) and to the (S)tarburst and
(N)ormal subsamples (dashed line) are given. Note that the same scaling is
used for each plot. 
}
\label{lx_lfir}
\end{figure}

$L_{X}$ correlates tightly with $L_{FIR}$ -- the scatter of the galaxies about
this relationship is remarkably small, the two main outliers being NGC~4258
and NGC~1291. That these two galaxies appear different from the norm is very
likely due to the facts that ({\rm i}) NGC~4258 is the most AGN-dominated
system in the survey, and is known to have excess diffuse X-ray emission
associated with two central AGN-produced `jets' (Paper~1; Pietsch \etal\ 1994),
and ({\rm ii}) NGC~1291 is of type Sa, and is the earliest type system in the
survey, with a very elliptical-like bulge and a consequently low far-infrared
luminosity (Bregman, Hogg \& Roberts 1995). The slope of the Fig.~5 (top) 
relationship
($L_{X} \propto L_{FIR}^{0.8}$) appears to be less than unity, and no
significant change in slope is seen between the normal and starburst galaxies
(see Table~\ref{table_slopes}).


The relationship between the purely {\em diffuse} X-ray
luminosity and the far-infrared luminosity (Fig.~5 [middle]) appears very
similar to that of $L_{X}$-$L_{FIR}$. $L_{X}^{\mbox{\small diff}}$
increases with $L_{FIR}$, and very little scatter is seen, apart again
from NGC~4258 and NGC~1291. However, here the 
slope is unity, $L_{X}^{\mbox{\small diff}} \propto L_{FIR}$. Again, no 
significant change is seen between the normal and the starburst galaxies.
The corresponding correlation coefficients are generally even higher than for 
$L_{X}$-$L_{FIR}$, especially for the starbursts.

If the total X-ray luminosity is increasing less rapidly than the far-infrared
luminosity, while the diffuse X-ray component keeps pace, then this
suggests that it must be the remaining component, the source X-ray component,
is lagging $L_{FIR}$. Indeed, as is seen in Fig.\,5, (bottom), and in
table~\ref{table_slopes}, the $L_{X}$(src)-$L_{FIR}$ slope is flatter
($L_{X}$(src)$ \propto L_{FIR}^{0.66}$). Note again, the lack of any 
significant difference between the normals and the starbursts. 

%
%

\subsection{X-ray$-$Optical luminosity relationships}
\label{sec_X-OPT}

Fig.~6 shows the relationships between the total (top), the 
diffuse (middle), and the source (bottom) X-ray 
luminosity, with the optical luminosity (a measure, remember, of the mass of 
the galaxy). 

\begin{figure}
\unitlength1.0cm
    \psfig{figure=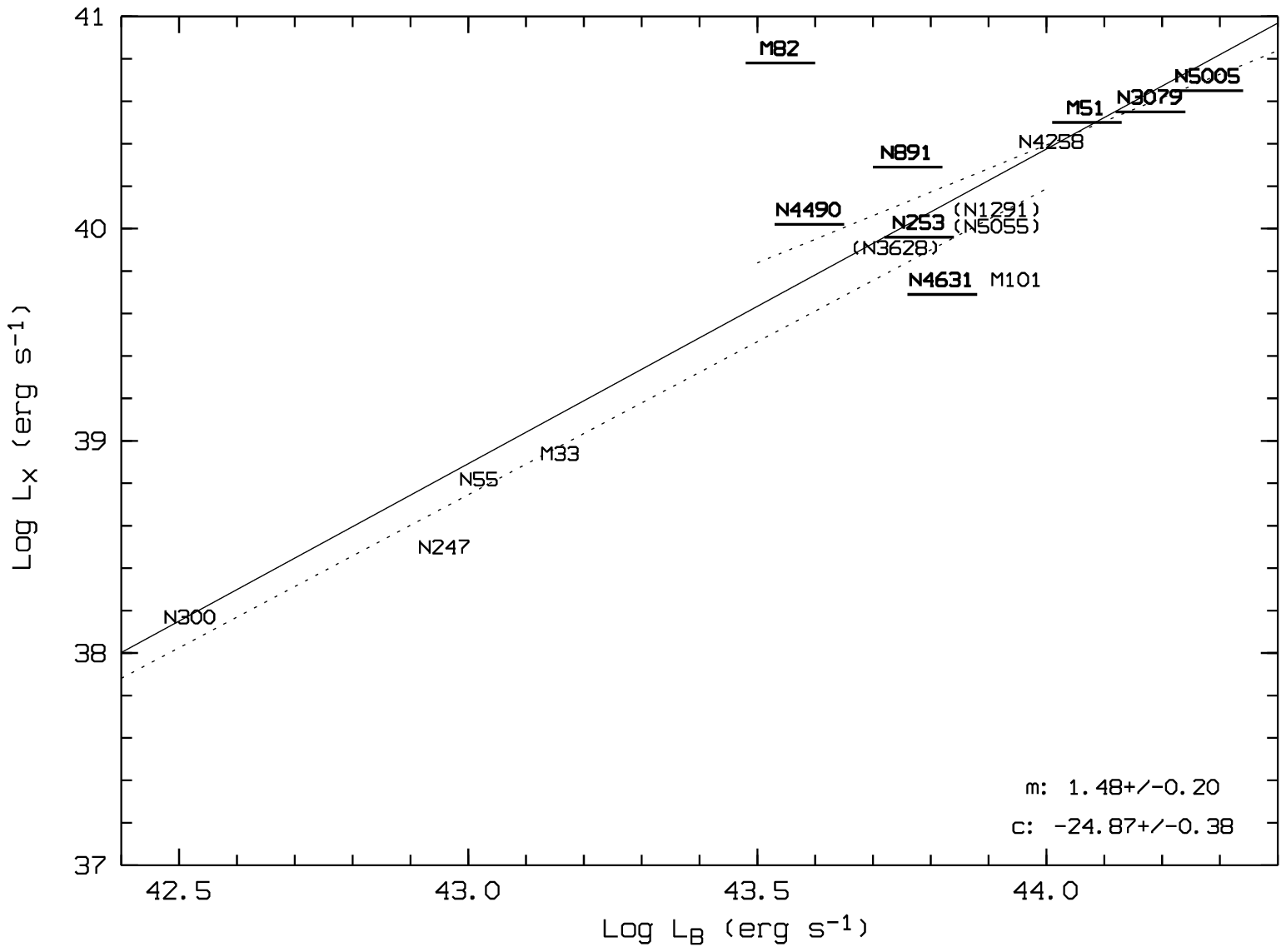,width=9.0cm,clip=}
    \psfig{figure=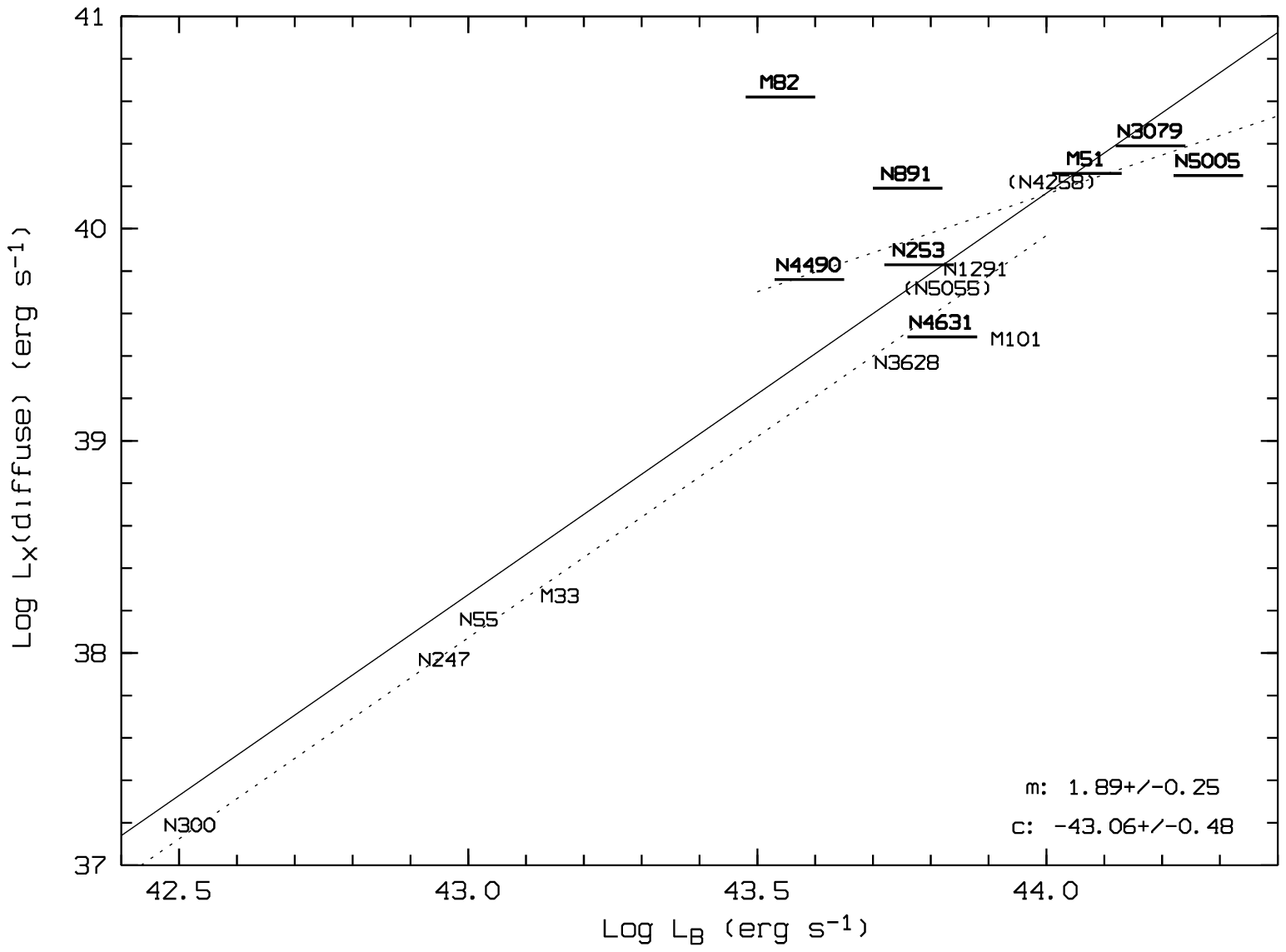,width=9.0cm,clip=}
    \psfig{figure=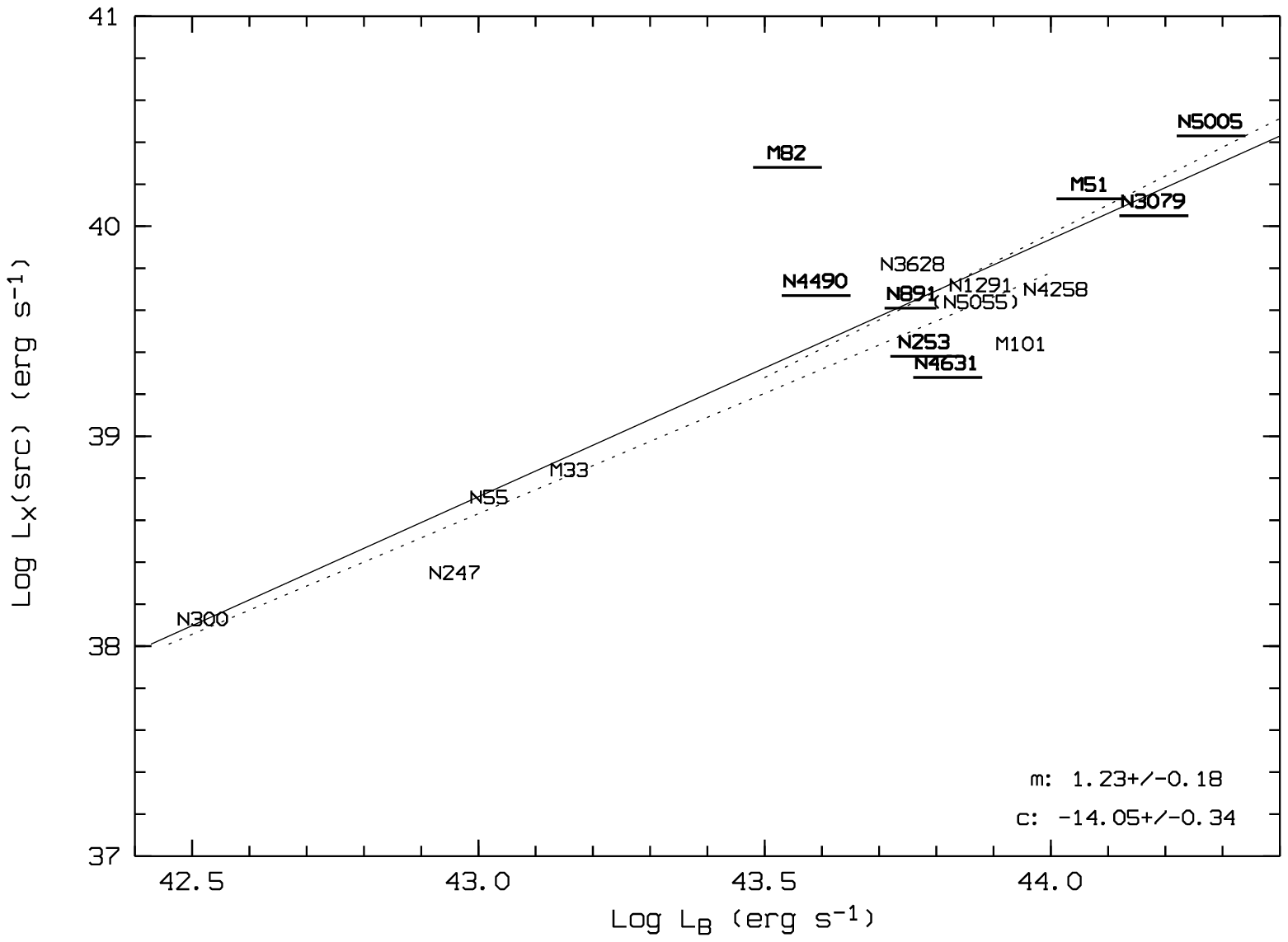,width=9.0cm,clip=}
\hfill \parbox[b]{9.0cm}
\caption{
\Ros\ (0.1$-$2.0\,keV) total (top), diffuse (middle) and source (bottom) 
X-ray luminosity versus optical luminosity for the survey sample. 
Starburst galaxies are marked in a bold font and 
underlined.
Regression fit lines to the total sample (bold line) and to the (S)tarburst and
(N)ormal subsamples (dashed line) are given. Note that the same scaling is
used for each plot. 
}
\end{figure}

As expected, $L_{X}$, $L_{X}^{\mbox {\small diff}}$ and $L_{X}^{\mbox {\small
src}}$ all increase with $L_{B}$. 
The total X-ray:optical slope is consistent with the recent {\em 
Einstein} work of Fabbiano \& Shapley (2000). 
Little scatter is seen at the low $L_{B}$
end of each plot, whereas a good deal of scatter is seen at higher $L_{B}$. 
This is evident in the correlation coefficients and regression fit results 
in Table~\ref{table_slopes}. In both the total and diffuse cases, the
normal galaxy subsamples are very well correlated, whereas stable 
fits 
(and even then, with a great deal of scatter) are only obtained for
the starburst subsample when M82 is removed. Note that while for the
starbursts, $L_{X}^{\mbox {\small diff}}$ appears well correlated with
$L_{FIR}$, for the normals it is very well correlated with $L_{B}$.
 
In the total and diffuse cases, 
several things can be seen. Firstly, for the whole sample, $L_{X}$ (and
$L_{X}^{\mbox {\small diff}}$) rise faster than $L_{B}$,
with slopes significantly greater than unity. Secondly, in both cases, the
total sample slopes are essentially identical to their respective normal
galaxy subset slopes. Thirdly, though no regression fits are obtainable (in
either case) for the full starburst galaxy subsets, when the `irregular' M82 is
removed from the analysis, the regression slopes obtained are significantly
lower, more like unity in both cases, and are significantly different from
both the total and normal subset slopes. 

For the source emission, the slope of $L_{X}^{\mbox {\small src}}$:$L_{B}$
is lower than for $L_{X}$:$L_{B}$ for the total sample, and the starburst and normal
subsamples -- all being nearly consistent with unity. In contrast
to the situation with diffuse emission, there is no indication
of any flattening of the relation for starbursts.

\subsection{The diffuse emission and hot gas properties}
\label{sec_hotgas}

Fig.\,7 shows the relationship between the X-ray luminosity,
$L_{X}^{\mbox{\small diff}}$, and
temperature, $T_{\mbox {\small diff}}$, of the diffuse emission.
The $L:T$ relation has been a topic of great interest in the study
of X-ray emission from galaxy clusters, since it is easy to
show that for hot gas with universal mean density, filling potentials
which scale in a self-similar way, $L_X$ is expected to scale
as $T^{1.5}\Lambda(T)$, where $\Lambda(T)$ is the temperature
dependence of the X-ray emissivity. In the case of bremsstrahlung,
$\Lambda(T)\propto T^{1/2}$, so one expects $L_X\propto T^2$. For gas with
$T<2$\,keV, such as we observe in spiral galaxies, $\Lambda(T)$ is very flat,
and therefore $L$ would rise more slowly that $T^2$.

Whilst self-similar potentials and universal gas densities may be
reasonable first order expectations in the case of clusters, this is
much less obviously the case for individual spiral galaxies. Here the
additional physics of heating and cooling which affect the baryons,
may be expected to modify both the potentials, and especially the gas 
fractions and gas density profiles. In addition, we know from its
morphology (Paper 1) that the diffuse X-ray emission in some systems
is dominated by a starburst-driven wind, whilst in others it arises
from a gravitationally bound corona, or from hot bubbles within the disc.

We should not be surprised, therefore, to see a large degree of scatter
in the relation shown in Fig.\,7. Nonetheless there is a marginally 
significant trend ($T_S=1.9$) with a logarithmic slope of $2.5\pm1.6$.
The correlation does not result from a scaling
with galaxy {\it size}, since $T_{\mbox {\small diff}}$ does not
correlate with $L_B$. Rather the relation
appears to arise principally from the fact that non-starbursts 
tend to have both lower temperatures and lower diffuse X-ray luminosities 
than starburst galaxies.

\begin{figure}
\unitlength1.0cm
    \psfig{figure=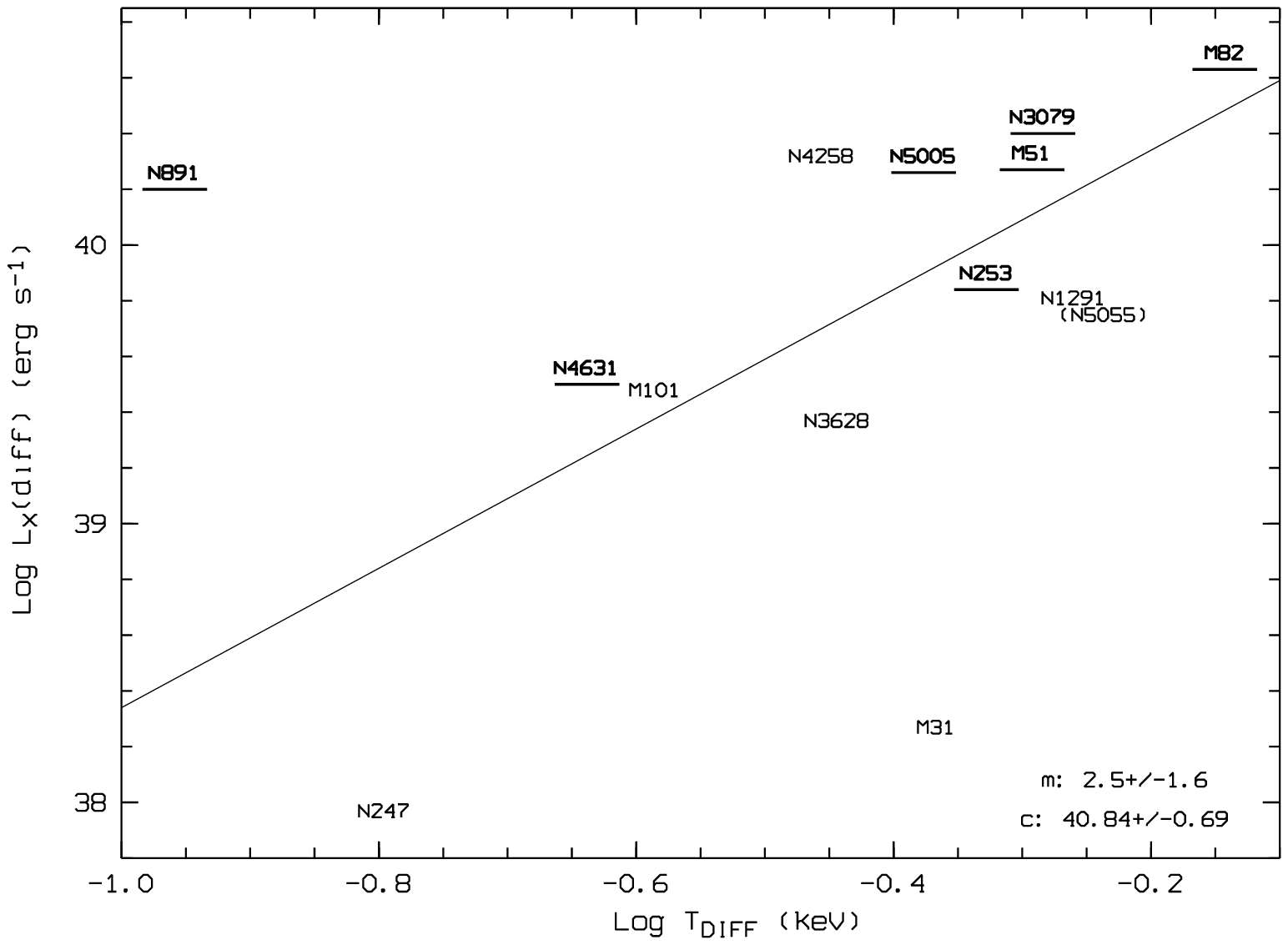,width=9.0cm,clip=}
\hfill \parbox[b]{9.0cm}
\caption{
Log-log plot of the diffuse emission temperature
$T_{\mbox {\small diff}}$ against the 
diffuse X-ray luminosity (0.1$-$2.0\,keV). 
}
\end{figure}

The relationship between the temperature and the
inferred mass of the hot gas is shown in Fig.~8. There is a tendency for  
more hot gas to be found in those systems exhibiting 
higher gas temperatures. No acceptable regression fits could be obtained, 
but the correlation coefficients [total sample:
0.57 ($T_{S}$ 2.4) starbursts: 0.87 ($T_{S}$ 3.9) normals 0.77 ($T_{S}$
1.7)] confirm what can be seen directly from Fig.~8 -- the correlation
arises principally from the starburst galaxies. In a similar vein
Fig.~9 shows that there is a relationship between the
inferred mass of the hot gas with the far-infrared luminosity, which
is especially marked amongst the starbursts. 

\begin{figure}
\unitlength1.0cm
    \psfig{figure=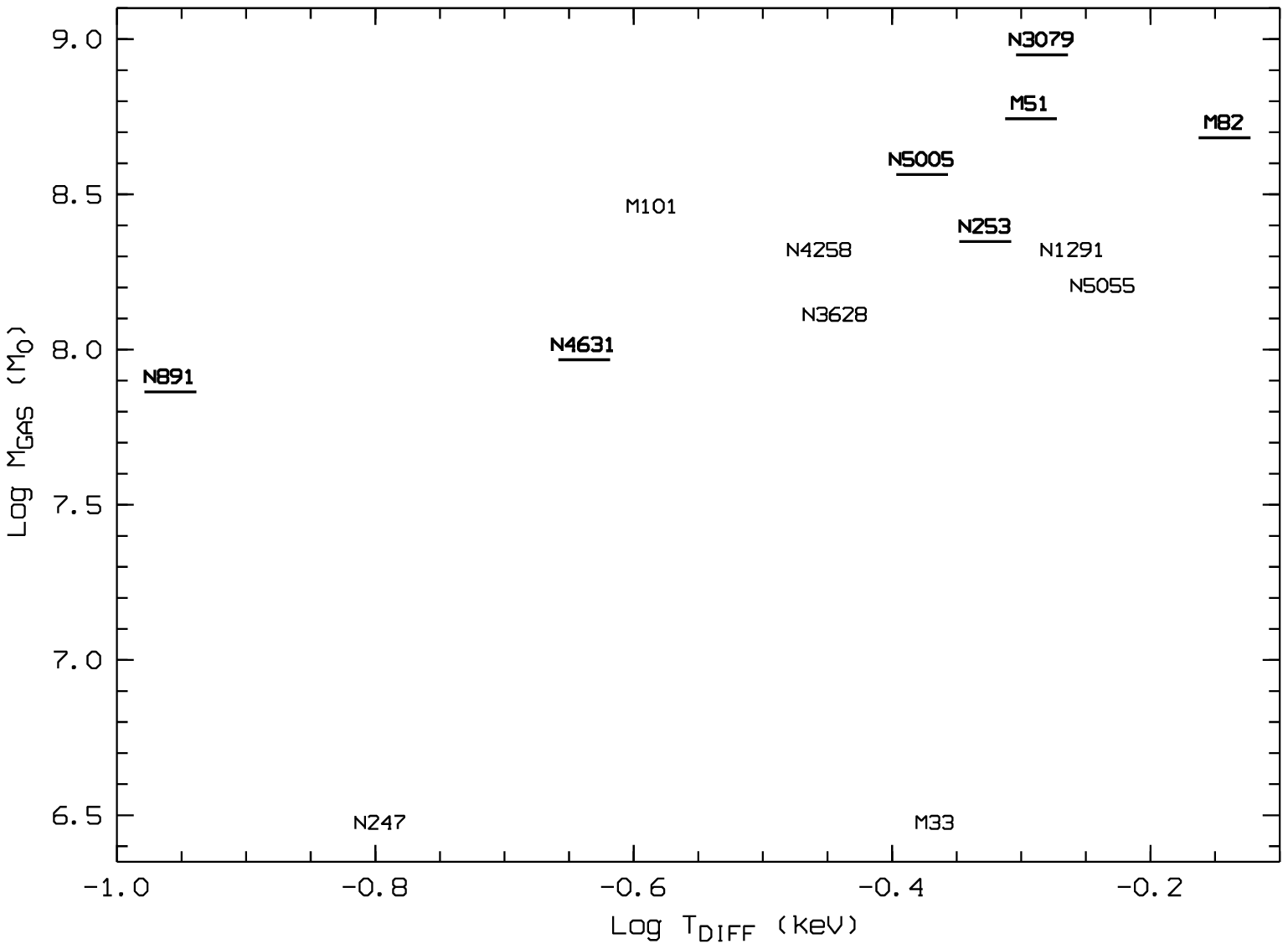,width=9.0cm,clip=}
\hfill \parbox[b]{9.0cm}
\caption{
The relationship between the diffuse gas mass and the diffuse gas X-ray
temperature for those members of the sample for which reliable measurements
have been obtained. 
}
\end{figure}

\begin{figure}
\unitlength1.0cm
    \psfig{figure=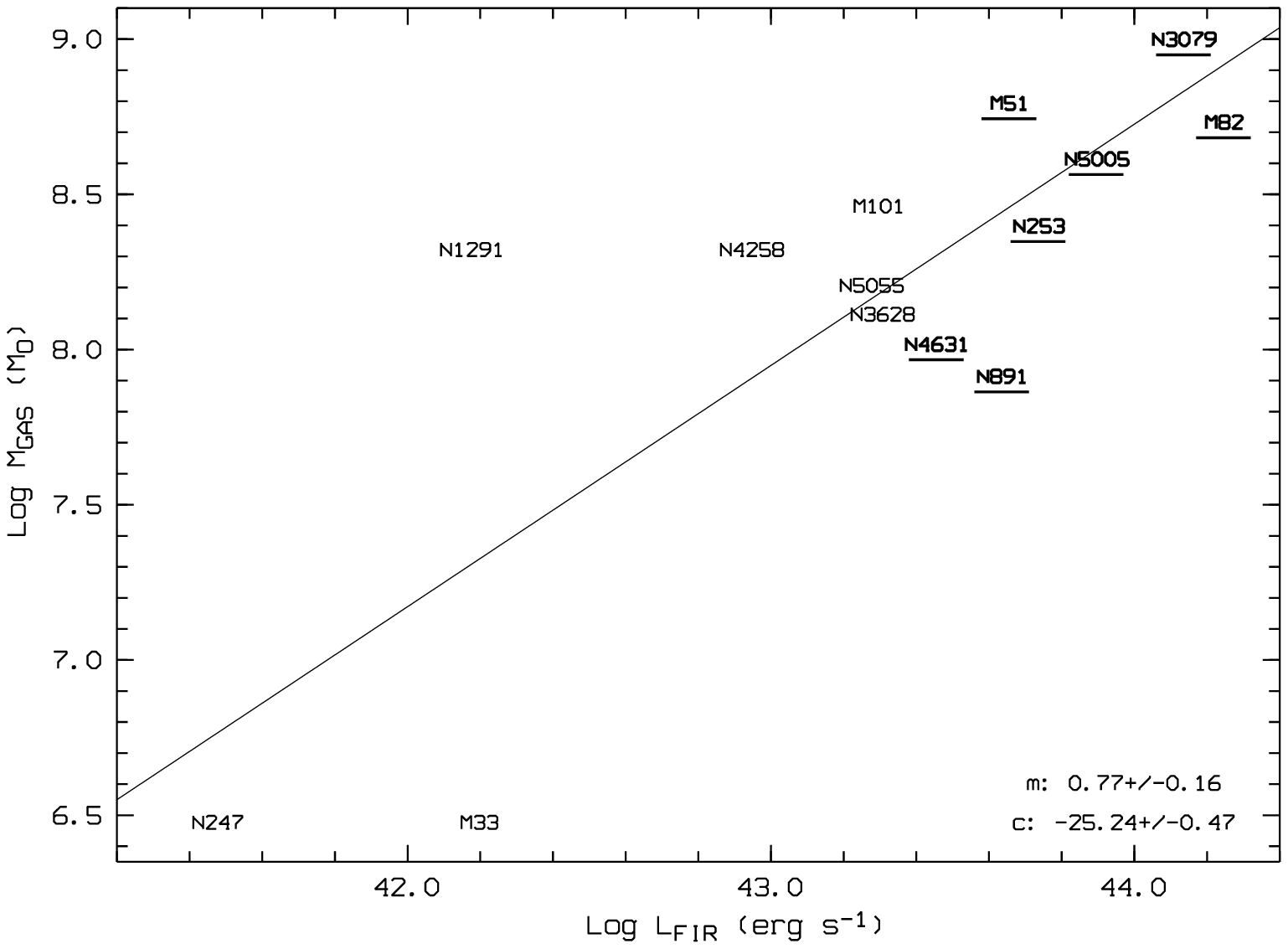,width=9.0cm,clip=}
\hfill \parbox[b]{9.0cm}
\caption{
The log-log relationship between the diffuse gas mass and the 
far-infrared luminosity for those members of the sample
for which reliable measurements have been obtained.
}
\end{figure}

\subsection{Variations in X-ray properties with Hubble galaxy type}
\label{X-sptype}

Fig.~10 shows the relationship between the diffuse gas temperature and
the Hubble type of the galaxy. Although we are dealing with small numbers
here, there is a suggestion, for the normal galaxies, that
the earlier Hubble types contain hotter gas (correlation
coefficient -0.54, $T_{S}$ -1.4), whereas for the starburst galaxies, the
later type systems are hotter (correlation coefficient 0.68, $T_{S}$ 2.1).

\begin{figure}
\unitlength1.0cm
    \psfig{figure=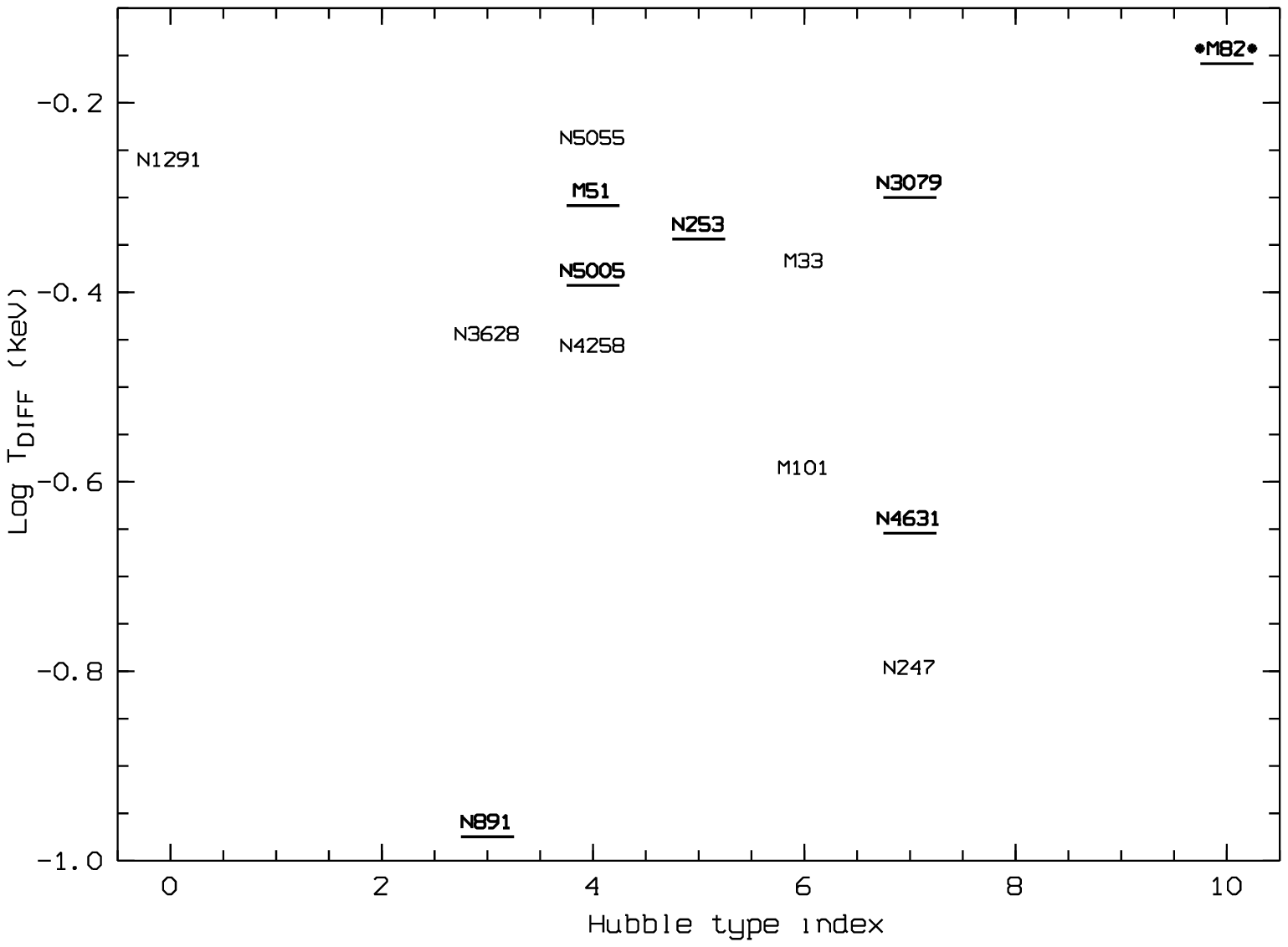,width=9.0cm,clip=}
\hfill \parbox[b]{9.0cm}
\caption{
The relationship between the diffuse gas temperature and the galaxy
Hubble galaxy type for those members of the sample for which reliable measurements
have been obtained (M82 has been marked with asterisks, as it is an irregular
galaxy). 
}
\end{figure}

The relationship between X-ray to optical luminosity ratio and Hubble type
is given in Fig.~11. Again for normal galaxies, a weak negative correlation is seen 
(\ie\ as we move towards earlier Hubble type, the X-ray to optical luminosity 
ratio appears to increase). However, nothing concrete can be said here 
regarding the starburst sample.

\begin{figure}
\unitlength1.0cm
    \psfig{figure=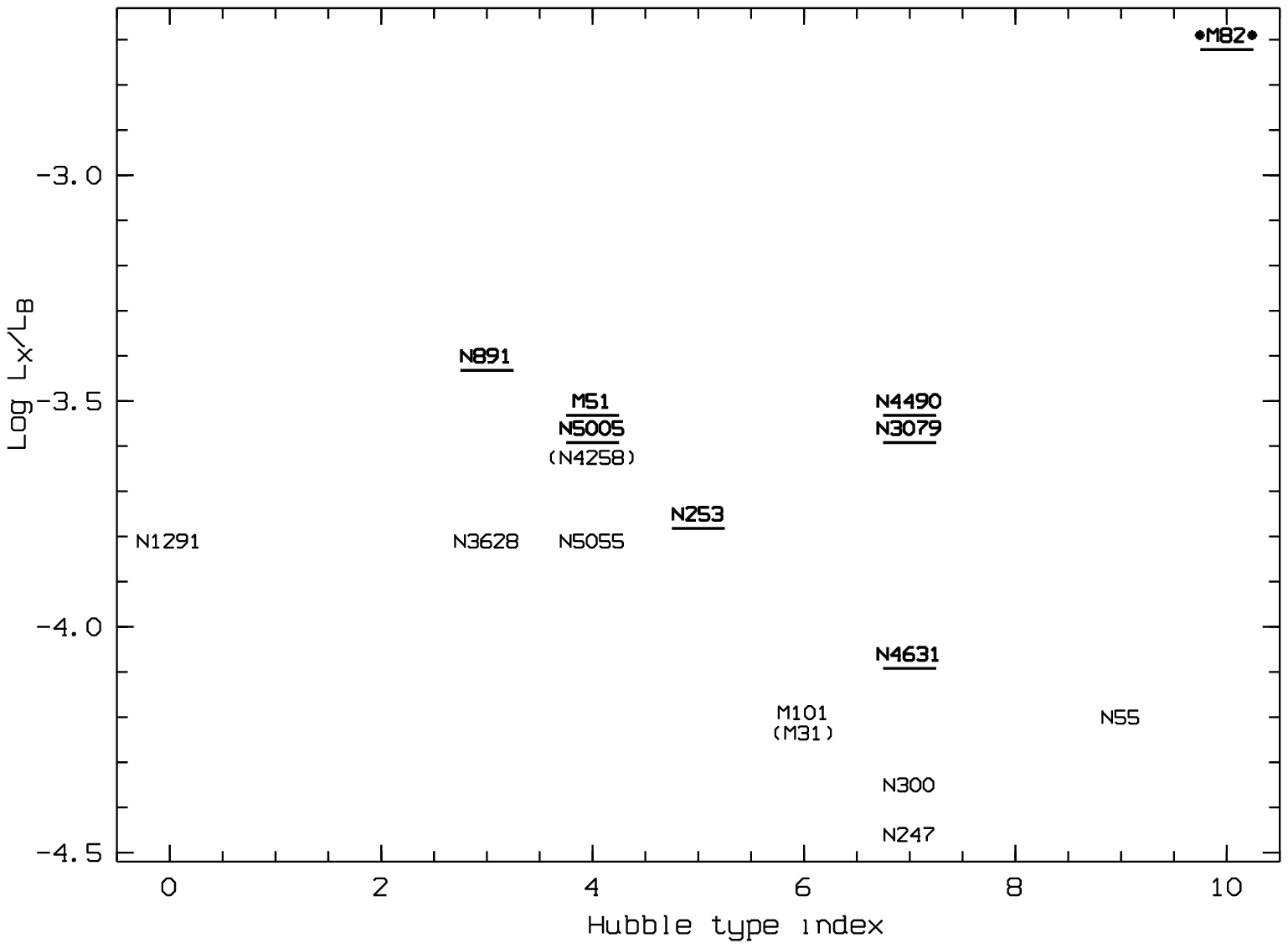,width=9.0cm,clip=}
\hfill \parbox[b]{9.0cm}
\caption{
The relationship between the X-ray to optical luminosity ratio and the galaxy
Hubble galaxy type for sample members (M82 has been marked with asterisks, as it is
an irregular galaxy). 
}
\end{figure}

Finally, it is worth noting the case of the earliest type system in the survey,
NGC~1291. This Sa-type galaxy appears to have a normal X-ray to optical
luminosity ratio (Fig.~11, Fig.~6), but a very high X-ray to far-infrared
luminosity ratio (Fig.~5). It is known that this system's X-ray luminosity and
radial surface brightness distribution are like that of elliptical galaxies,
and that, like ellipticals, the $\log{L_{FIR}}-\log{L_{B}}$ ratio is small
(Bregman \etal\ 1995). This can be seen in Figs.~1 \& 3. Throughout this
paper, NGC~1291 is seen to appear elliptical-like in that its far-infrared
luminosity is reduced in comparison with its X-ray and optical luminosity,
whereas the next earliest galaxies, NGC~891 and NGC~3628, appear like typical
spirals.

\section{Discussion}
\label{sec_discussion}

In the previous section, we presented the main results from a  
statistical analysis of the galaxy sample. In discussing these results, 
we concentrate here on the parameters which have not generally been available from
earlier studies. In particular, we have been able to 
split each system's X-ray emission into a source component and 
a diffuse, gaseous component, and also to investigate the spectral properties
of the hot gas. Our `new' X-ray parameters are therefore the total, source and diffuse 
X-ray luminosities, and the diffuse emission temperatures (together with 
related parameters, such as gas mass \etc). 

\subsection{X-ray emission as a function of mass and activity}
\label{sec_disc_massact}

The broad picture which emerges from the results presented in
sect.~\ref{sec_X-massact} is that
for starburst galaxies, X-ray properties scale with activity rather than mass. 
For the normal galaxies, in contrast, X-ray properties are strongly correlated with 
mass, and not with activity. 

Large normal galaxies, irrespective of activity, appear to possess more diffuse
emission (here assumed to be primarily hot gas - see Section~2 and Paper~1), 
than smaller galaxies. This is not a simple a matter of scaling with size,
since the diffuse X-ray luminosity {\it per unit galaxy mass}, and the diffuse
emission fraction both increase with $L_B$. Since the luminosity of hot gas
scales as $n_e^2V$, a simple size scaling leads to $L_X\propto L_B$, if mean density
and gas fraction do not change with galaxy mass. It appears therefore that large
galaxies are able to heat or retain a larger fraction of their gas than do smaller
galaxies, or they compress it to higher mean densities.
One plausible picture is that hot gas from active star formation regions is
able to escape more readily from the shallower potential wells of small galaxies,
systematically lowering their hot gas fraction.

In the case of starburst galaxies, the fraction of diffuse emission
(\ie\ hot gas) is primarily determined by activity rather than galaxy
mass. This suggests that the influence of the galaxy potential,
important for the normal galaxies above, is no longer as
important. The activity now dominates, having reached such a level
that whether the galactic potential is large or small has little
effect.

This is all consistent with the idea, supported by the morphology of
the emission (Paper 1), and by simulations and observations at other
wavelengths, that in normal galaxies one sees bound `coronal' gas,
whilst in starburst galaxies the bulk of the hot gas is a freely
escaping hot wind.


Turning to the source component of the X-ray emission, this is found to scale
approximately linearly with $L_B$ for both normal and starburst galaxies. 
This result is readily understood if the discrete source complement scales
with the overall stellar mass, and suggests
that the separation of source and diffuse contributions in our sample has
been successfully achieved. Although the principal scaling is with galaxy
mass, there is also a trend of increasing $L_{X}^{\mbox {\small src}}/L_B$
with galaxy activity (Fig.3). Since we believe that source confusion does not
have a major effect on our separation of diffuse and source emission (section~2),
the implication is that there is a significant rise in the incidence of
discrete X-ray sources in systems which are actively forming stars -- i.e.
these galaxies contain a significant contribution from short-lived sources,
such as high mass X-ray binaries [HMXB] or young SNR. In contrast, we know
that the source population in the Milky Way is dominated by the older
low mass X-ray binary (LMXB) population.

\subsection{The effect of star formation rate}
\label{sec_disc_multiwave}

We saw in Sect.~\ref{sec_X-FIR} that the {\em diffuse} 
X-ray emission rises proportionally with the
far-infrared luminosity (\ie\ SFR), whilst {\em source} X-ray luminosity rises 
less steeply. Combining diffuse and source contributions gives the observed 
$L_{X}$(total)$\propto L_{FIR}^{0.8}$ relationship. There is therefore a shift
in the balance between source and diffuse X-ray luminosity 
as the SFR increases. This is quite easy to
explain in terms of more of the high-SFR sources being diffuse in nature -
supernova remnants and collections thereof, superbubbles, supernova-produced
fountains, halos and coronae of hot gas, and galactic winds. Furthermore, we
would not expect the {\em source} X-ray luminosity to keep pace with $L_{FIR}$
(\ie\ the current SFR), as it is dominated by an older component (mostly 
LMXBs). The above explanation is strengthened by the 
fact that the diffuse X-ray component is seen to be more tightly 
correlated with $L_{FIR}$ than the source component .

We also found that there appears to be no significant change in the X-ray:FIR 
slope (in any component, whether total, diffuse or source) as we go from
(low-activity) normal galaxies to (high-activity) starburst galaxies.
The lack of any change in the response of the diffuse luminosity to SFR
is interesting and surprising, given that we believe that the transition
from an essentially hydrostatic hot corona to a wind occurs within this range.
Furthermore, Read \& Ponman (1998) found that, in even higher activity 
{\em interacting} and {\em merging} systems, a change in the 
X-ray:FIR slope {\em is} seen; the logarithmic X-ray:FIR slope is seen to
drop significantly to around 0.3$-$0.4. It was suggested that this 
relative deficit in X-ray emission might be due to much of the 
input energy from supernovae and 
stellar winds in these very high-activity systems, being converted
into the kinetic energy of the huge plumes of ejected gas seen in these 
systems, rather than contributing to the X-ray emission.
In contrast, the high-activity systems in the present sample 
do not behave like these merging systems, even though in several cases 
(M82, NGC~253, NGC~3079 \etc), large plumes of hot gas do exist. 

The work of Fabbiano \& Shapley (2000), who analyse the results of a 
multiwavelength statistical study of the {\em Einstein} sample of spiral 
and irregular galaxies, is consistent with the above results. These authors 
derive X-ray:FIR slopes very similar to our present study. They 
also see a suggestion of the X-ray:FIR slope flattening, though in their
data it is still 
consistent with a constant slope. This is of interest as the Fabbiano \& 
Shapley (2000) study did include very high-activity galaxies, galaxies where, 
as in the Read \& Ponman (1998) study, kinetic energy losses had become
important.

It is well established (e.g. Read \& Ponman 1995) that the X-ray luminosity
of late-type galaxies is only a small fraction of their total supernova
luminosity in all cases. The bulk of the supernova energy is apparently
radiated in the FIR. It appears, however, that an approximately constant 
fraction of the energy released during the life and death of stars is
radiated by hot gas in both normal and in moderate starburst galaxies, but
the massive kinetic energy losses suffered in the ultraluminous outbursts
associated with galaxy merging, reduces the fraction of the energy
available for radiating in the X-ray band.


\subsection{Diffuse components and their spectral properties}
\label{sec_disc_hotgas}

Before discussing the spectral results relating to the diffuse components, we should
consider first what these diffuse components might be. As discussed in Paper~1 and again
in Section~2, we believe the diffuse components observed to be due mainly to hot gas
produced within these systems. It is believed, both through the literature and through
the results discussed in Sect.~\ref{sec_disc_massact}, that two forms of hot X-ray
emitting gas can exist in the halos of these spiral galaxies.

Some edge-on systems within our sample (\eg NGC~891 and NGC~4631; see Paper~1)
exhibit what is believed to be {\em coronal} gas (Bregman \& Pildis 1994; Wang
\etal\ 1995). This cool (1$-$3\,million~K) gas, is thought to bubble  out of
the plane of the galaxy through galactic `chimneys' and `fountains' (formed by
localized high-activity star-forming regions within the disc), then to fall
ballistically back to the plane (Norman \& Ikeuchi 1989). This X-ray emitting
gas is thought to be in approximate hydrostatic equilibrium and occupies
almost the entire volume. Its filling factor $\eta$ is therefore thought to be
close to unity and this has been backed up with numerical multi-gas-phase
simulations (\eg\ Rosen \& Bregman 1995). 

Other edge-on systems within our sample (\eg\ NGC~253, M82 and NGC~3079; see
Paper~1) exhibit what is believed to be an outflowing {\em wind}.
Many of these structures appear to be very extended, having temperatures of
$T\sim4-8\times 10^6$\, K. The structure of the gas is not well known. On the
one hand, if the gas filling factor $\eta$ is close to unity, then these
features could be due to plumes of gas flowing out from the galaxy. On the
other hand, simulations (Suchkov \etal\ 1994, Strickland \& Stevens 2000)
and observations (Heckman \etal\ 1993, Strickland \etal\ 2000)
suggest that in these systems, most of the
X-ray emission is due to small `clouds', shock-heated by a fast
($\sim3000$\,km s$^{-1}$), hot ($\gg10^{7}$\,K) wind, produced through the
efficient thermalization of massive stellar winds and supernovae ejecta within
a starburst nucleus. This wind is believed to be
orders of magnitude less luminous in the \Ros\ band (Suchkov \etal\ 1994) . In
this scenario, the filling factor $\eta$ of the \Ros\ X-ray emitting gas is
very much smaller than unity.

Note that if this is the case, the gas masses inferred for the wind systems
under the assumption $\eta=1$ may be seriously overestimated. The trends in
$M_{\rm gas}$ from quiescent to more active galaxies, plotted in Figs.~8 \& 9,
may therefore be misleading. It is likely that the gas mass estimates
for coronal systems are reasonably representative, whilst those for
winds may be too high.

Considering edge-on systems, in which the morphology of the
hot gas can be most clearly seen, we note that the
diffuse emission in the `coronal' systems (NGC~4631 and NGC~891) appear
cooler than those seen within the known `wind' systems
(NGC~253, M82 and NGC~3079). As suggested in Paper~1, this temperature
diagnostic suggests that the diffuse emission structures seen in NGC~5005 and
NGC~5055 are more likely to be due to winds than coronae.

Diffuse structures are also seen in many of the face-on systems (\eg M33, M51,
M101). We have no idea on geometrical grounds, how far these diffuse
emission features extend above the plane. As discussed in Paper~1, the
diffuse emission seen in M33 is low-power, and only occupies the central area
of the disc. In M51 and M101 however, the diffuse features observed are both
large, powerful, and fairly uniform. Their respective positions in Fig.\,8
suggest, assuming the coronal/wind argument described above, that the diffuse
emission in M51 is dominated by galactic wind emission, whereas in M101, it is
dominated by coronal emission.

Presumably, there is nothing to stop both coronae and winds existing
simultaneously in a given galaxy, and this may have been observed in the
merging galaxy pair NGC~4038/9 (Read, Ponman \& Wolstencroft 1995).



\section{Conclusions}
\label{sec_conclusions}


We have reported the results of a statistical analysis of a
study of 17 spiral galaxies with the 
\Ros\ PSPC. Contributions to the X-ray emission from discrete sources 
and diffuse emission have been carefully separated, allowing us to
probe both the hot gas and X-ray sources.
The most interesting conclusions from this work are:

%

\begin{itemize}

\item In general, the X-ray properties of normal galaxies are governed by the 
galaxy size, whereas the X-ray properties of starburst galaxies are governed 
by the galaxy activity. 

\item Larger normal galaxies contain more hot gas per 
unit mass than smaller ones. We interpret this as resulting from better retention
of hot gas in the deeper potential wells of larger galaxies.

\item In contrast, the diffuse emission per unit mass from starburst galaxies 
scales with activity rather than galaxy size. This is consistent with the idea
that the hot gas in these systems is substantially unbound, and its density
is therefore determined by the rate of hot gas production rather than by 
the retaining potential.

\item The X-ray emission per unit mass from discrete sources 
does not depend on the total galaxy mass, but does rise somewhat with galaxy
activity. This implies that in the more active systems, young objects such as
HMXB or compact SNR make a significant contribution to the X-ray source population.

\item Fits to $L_{X}$:$L_{B}$ and $L_{X}$:$L_{FIR}$ trends
agree well with previous studies. Extracting the behaviour of the hot gas, we find
a remarkably tight linear scaling with $L_{FIR}$ which applies across the
activity range covered by this study, though it appears to be broken in
ultraluminous mergers. The reason for the continuity in this relation
across the transition from coronal to wind emission is far from clear, and
the authors hope that future hydrodynamical modelling work will address this issue.

\item Gas in the `coronal' systems appears to be cooler than that
seen in the `wind' systems. This may help to diagnose the state of hot gas
in face-on systems.

\end{itemize}

\section*{Acknowledgements}

AMR acknowledges the receipt of a Royal Society fellowship during the 
early stages of this work, and of support from the MPE ROSAT group during 
the later stages. 
This research has made use of the
SIMBAD database operated at CDS, Strasbourg. The ROSAT project is supported
by the German Bundesministerium f\"ur Bildung, Wissenschaft, Forschung
und Technologie (BMBF/DLR) and the Max-Planck-Gesellschaft (MPG).

\end{document}